\documentclass[a4paper,11pt]{article}
\usepackage{aaskaiid}
\usepackage{orcidlink}
\usepackage{graphicx}
\usepackage{newtxtext,newtxmath}
\usepackage{main}

\usepackage[dvipsnames,svgnames,x11names]{xcolor}

\usepackage{enumitem}

\title{The SKA View of the Sunyaev-Zeldovich Effect\newline from Massive Cosmic Halos}
\ShortTitle{SKA View of the SZ Effect}

\author[1]{Yvette C. Perrott\orcidlink{0000-0002-6255-8240}}
\ShortName{Yvette Perrott et al.} 
\author[2,3]{Luca Di Mascolo\orcidlink{0000-0003-3586-4485}}
\author[3]{Rémi Adam\orcidlink{0009-0000-5380-1109}}
\author[3]{Chiara Ferrari\orcidlink{0000-0002-2917-9759}}
\author[4]{Keith J. B. Grainge\orcidlink{0000-0002-6780-1406}}
\author[5]{Matthias Hoeft\orcidlink{0000-0001-5571-1369}}
\author[6]{Mamta Pandey-Pommier\orcidlink{0000-0001-5829-1099}}

\affiliation[1]{School of Chemical and Physical Sciences, Victoria University of Wellington, PO Box 600, Wellington, New Zealand}
\affiliation[2]{Kapteyn Astronomical Institute, University of Groningen, Landleven 12, 9747 AD, Groningen, The Netherlands}
\affiliation[3]{Universit\'e C\^ote d'Azur, Observatoire de la C\^ote d'Azur, CNRS, Laboratoire Lagrange, France}
\affiliation[4]{Jodrell Bank Centre for Astrophysics, Department of Physics \& Astronomy, The University of Manchester, Manchester M13 9PL, UK}
\affiliation[5]{Th\"uringer Landessternwarte, Sternwarte 5, 07778 Tautenburg, Germany}
\affiliation[6]{Pole Scientific, University Catholic of Lyon, Campus Saint-Paul, 10 place des Archives 69288, Lyon Cedex 02, France}

\emailAdd{yvette.perrott@vuw.ac.nz}
\emailAdd{l.di.mascolo@astro.rug.nl}
\emailAdd{remi.adam@oca.eu}
\emailAdd{chiara.ferrari@oca.eu}
\emailAdd{keith.grainge@manchester.ac.uk}
\emailAdd{hoeft@tls-tautenburg.de}
\emailAdd{mamtapommier@gmail.com}

\abstract{The thermal intracluster medium (ICM) can be observed via its interaction with Cosmic Microwave Background photons, known as the Sunyaev-Zeldovich (SZ) effect.  This effect produces an observable signal at radio to sub-mm wavelengths which probes the pressure of the ICM.  The SKA will be sensitive to the thermal SZ effect in its highest frequency band, 5b.  In this Chapter, we show that the SKA will provide a high-resolution, high-sensitivity view of the thermal SZ effect, allowing detailed observations of pressure substructures in clusters while retaining sensitivity to the large-scale global ICM emission.}

\begin{document}

\maketitle

\section{Introduction}

Galaxy clusters, massive gravitationally bound structures existing at the intersection of filaments in the cosmic web, are excellent probes of both astrophysics and cosmology thanks to their sensitivity to underlying cosmological parameters \citep{2011ARA&A..49..409A} along with the interactions of their constituent galaxies, intracluster medium and dark matter \citep{2012ARA&A..50..353K}. The intracluster medium (ICM), a hot ($\sim$ keV) plasma pervading the galaxy cluster structure interacts both thermally and dynamically with its environment.  It provides a sensitive probe of the dark-matter-dominated total mass \citep{2019SSRv..215...25P} and an observationally inexpensive but robust method for detecting clusters. The thermal ICM is detected in X-rays, where it emits via Bremsstrahlung and line emission processes, and in the radio to sub-millimetre waveband, where inverse Compton scattering of Cosmic Microwave Background (CMB) photons creates a distortion of the CMB background emission known as the Sunyaev-Zeldovich (SZ) effect (\citealt{1970Ap&SS...7....3S}; for a recent review, see \citealt{2019SSRv..215...17M}).  Although there are multiple variations of the SZ effect (kinetic, relativistic, etc), in this Chapter we focus on the thermal SZ (tSZ) effect in the non-relativistic approximation.

The strength of the tSZ signal is proportional to the line-of-sight integral of the electron pressure, whereas the X-ray signal roughly scales with the integral of the square of the electron density along the line of sight \citep{Sarazin1985}. This difference makes high-resolution tSZ observations a valuable complement to X-ray measurements, as tSZ observations:

\begin{enumerate}
    \item are more sensitive to the outskirts of clusters where the squared density dependence of X-ray makes the signal faint;
    \item are less sensitive to clumps of cooler gas along the line of sight, which produce biases in the X-ray measurements;
    \item in combination with X-ray data allow for the extraction of temperature information with less uncertainty than constraints from fitting to X-ray spectra.
\end{enumerate}

Moreover, the surface brightness of the tSZ signal is independent of redshift, in contrast to X-ray flux, which diminishes with redshift as $(1+z)^4$ due to cosmological dimming \citep{Carlstrom2002}. This makes the tSZ effect a powerful tool to detect high-redshift clusters and possibly characterize their dynamical states, provided the sensitivity to the low surface brightness signal and the angular resolution to resolve cluster substructure are sufficient.

SZ-based cluster catalogues derived from instruments such as \emph{Planck} (e.g.\ \citealt{Planck2016}), the South Pole Telescope (e.g.\ \citealt{Bleem2020}) and the Atacama Cosmology Telescope (e.g.\ \citealt{Hilton2021}) have been used to derive constraints on standard cosmological parameters such as $\Omega_\mathrm{m}$ and $\sigma_8$ as well as the dark energy equation of state parameter $w$, and to put upper limits on neutrino masses (e.g.\ \citealt{2013JCAP...07..008H,Planck2016b,2024PhRvD.110h3510B}).  However, the instruments that have produced these SZ catalogues have provided relatively low angular resolution views at $\sim$ arcmin resolution, suitable for measuring global (integrated) ICM properties rather than resolving ICM substructure.  In comparison, X-ray observations with instruments such as \emph{Chandra} have allowed detailed views of ICM substructure at $\sim$ arcsec resolution, enabling the detailed study of physical processes, e.g. those associated with possible acceleration mechanisms of intracluster cosmic rays.

Recently, updated instruments have begun to achieve higher resolution SZ effect observations.  These include both single-dish instruments, such as MUSTANG\slash MUSTANG-2 on the Green Bank Telescope (e.g. \citealt{2011A&A...534L..12F}, \citealt{2022A&A...667L...6O}, \citealt{Romero2023}) and the NIKA\slash NIKA2 camera on the IRAM telescope (e.g. \citealt{2018A&A...615A.112R}, \citealt{2018A&A...614A.118A,2024A&A...684A..18A}), and interferometers such as ALMA (e.g. \citealt{Kitayama2016}, \citealt{2016ApJ...829L..23B}, \citealt{2023Natur.615..809D}, \citealt{2024A&A...689A..41V}) and NOEMA (e.g. \citealt{2025A&A...702A.275M}).  These higher-resolution observations have begun resolving substructures, such as disrupted pressure morphologies and asymmetries in the distribution of the ICM, particularly in disturbed or merging systems, thereby providing insight into the physical processes governing cluster evolution.  However, there are trade-offs between higher frequency (better angular resolution and stronger tSZ signal, but also higher atmospheric noise), and interferometric observations (which resolve out large scale emission) vs single-dish observations (which suffer more from systematics, atmospheric contamination and generally have lower angular resolution).  SKA-MID, observing in Band 5b (8.3-15.4 GHz), will fill a unique niche amongst these instruments, and provide competitive high-resolution views in reasonable observation times, as we will show in this Chapter.

The next decade will revolutionize our ability to probe high-redshift ($z > 0.6$) galaxy cluster formation, thanks to new X-ray instruments (e.g. NewAthena; \citealt{newathena25}) and sub-mm observatories (Simons Observatory; \citealt{SO2019,SOLAT2025}). SKA-MID, with its high sensitivity and spatial resolution, offers unique capabilities for studying cluster dynamics and imaging substructures through tSZ. Using Band 5’s shorter baselines, SKA will detect the redshift-independent tSZ signal and observe clusters in a single pointing. SKA will map pressure profiles at small-to-intermediate radii and, utilizing its longer baselines, reveal detailed substructures like AGN bubbles and pressure discontinuities. 

Thanks to lower-frequency radio observations, we also know that the ICM hosts a non-thermal component, as relativistic electrons gyrating in the intracluster magnetic fields give rise to diffuse synchrotron emission on Mpc scales \citep{2008SSRv..134...93F,vanWeeren2019}. The presence of these relativistic electrons is thought to be primarily linked to turbulence and shocks generated during cluster mergers \citep{Brunetti2014}.  Combining SKA-MID Band 5 with Bands 1 and 2 (and possibly SKA-LOW) will allow the investigation of diffuse non-thermal emission from the ICM and surrounding filamentary structures, as well as galaxy evolution in clusters.
This will significantly advance cluster science by providing deeper insights into the physical processes governing cluster evolution, gas dynamics, and the impact of extreme cluster environments on galaxy evolution. 

In this Chapter, we present an exploration of the SZ observations that will be possible with the SKA using simulations and mock analysis procedures.  In Sect.~\ref{sec:model}, we describe the simulation setup employed throughout the studies presented in Sect.~\ref{sec:sim}. We begin by exploring the mass-redshift space that will be accessible with the SKA (Sect.~\ref{sec:sim:massz}) and investigating the prospects for constraining pressure profiles (Sect.~\ref{sec:sim:profiles}).  Then, we assess the possibilities for observing cluster substructure such as turbulent fluctuations (Sect.~\ref{sec:sim:perturb}) and ICM shocks (Sect.~\ref{sec:sim:shock}).  Finally, in Sect.~\ref{sec:discussion}, we compare potential SKA configurations, the effects of observing at higher frequency bands (which have been considered for the SKA) and discuss synergies with other instruments.

\section{Modelling framework}\label{sec:model}

\subsection{Simulations and mock observations}
All the simulations and mock observations presented in this work are generated using \texttt{SKASZ}\footnote{\url{https://github.com/lucadimascolo/skasz}}, a custom pipeline integrating several submodules from the RASCIL\footnote{\url{https://gitlab.com/ska-telescope/external/rascil-main}} and SKA Science Data Processor\footnote{\url{https://gitlab.com/ska-telescope/sdp}} (SKA SDP; \citealt{Broekema2015}, \citealt{Farnes2018}) libraries. The baseline-level sensitivity estimates are obtained on-the-fly using a dedicated porting of the \texttt{python} backend of the official SKA Sensitivity Calculator\footnote{\url{https://gitlab.com/ska-telescope/ost/ska-ost-senscalc}}. We refer to the documentation of the individual packages and to the \texttt{SKASZ} code base for further details.

\subsection{SKA parameters}

In Table~\ref{tab:ska_params} we collect together the instrumental parameters we use for the simulations.  We note that the current baseline design for AA4 does not include Band 5 receivers on the MeerKAT antennas so we exclude them except where explicitly mentioned (this will be explored further in Section~\ref{sec:discussion:meerkat}).  We assume the two tunable 2500\,MHz bands will be placed within the 5b frequency range to avoid the expected radio frequency interference at 10.8 -- 12.8\,GHz.  We use the untapered primary beam model provided in the SKA SDP library, and quote the full width at half maximum (FWHM) given by a Gaussian fit to the central lobe for reference.

\begin{figure}[!h]
    \centering
    \includegraphics[width=\linewidth]{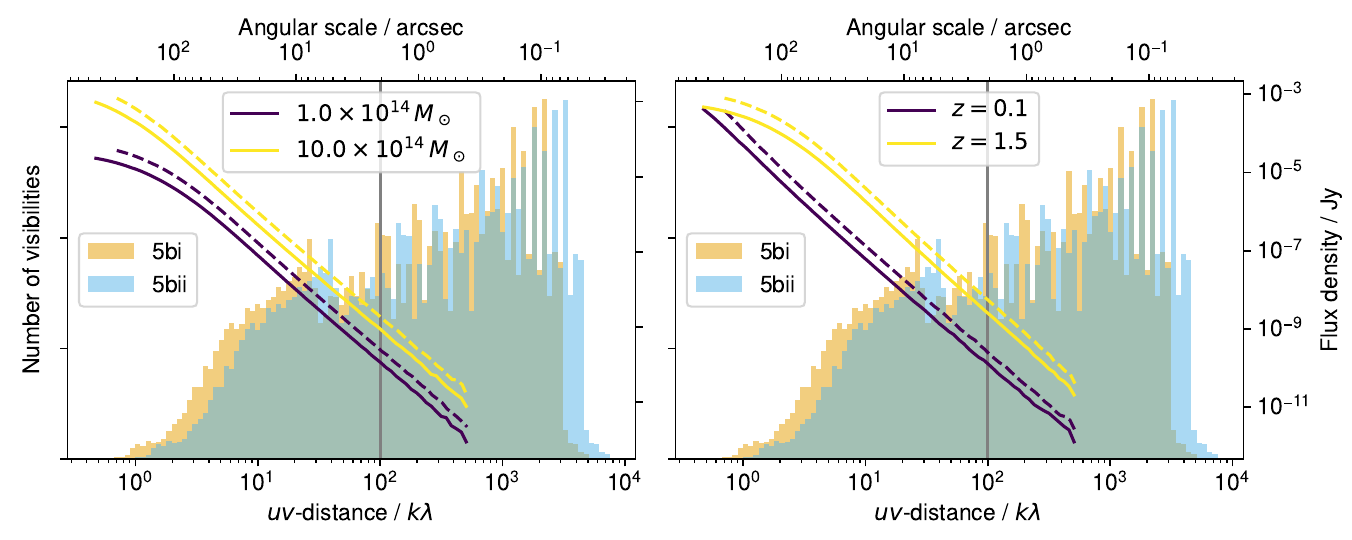}
    \caption{Representative examples of cluster profiles in $uv$-space (lines, right-hand axes) compared to the SKA-MID baseline distribution for a 10 hour observation (AA4, no MeerKAT; histograms, left-hand axes).  `5bi' and `5bii' refer to the lower and upper frequency bands within 5b, respectively.  Solid (dashed) lines show cluster signal profiles in 5bi (5bii).  In the left-hand plot, $z=0.5$ and mass is indicated in the legend; in the right-hand plot $M_{500} = 5 \times 10^{14} M_\odot$ and redshift is indicated in the legend.  The vertical line indicates the $uv$-range used for analyses of the SZ effect.  In real observations, the longer baselines would be used for removal of contaminating radio sources.}
    \label{fig:uvamps_A10}
\end{figure}

Since the large-scale cluster emission is only significant on baselines with length $\lessapprox 10^5 \lambda$ (see Figure~\ref{fig:uvamps_A10}), we cut off our simulated data at this $uv$-distance for analysis purposes.  In a real observation, the longer-baseline data would be used to constrain and subtract foreground and background radio sources.  Due to the very large number of these longer baselines, sensitivity to these confusing sources will be very high and therefore removal will be very accurate.

\begin{table}
    \centering
    \begin{tabular}{||l|lll|}
        SKA array & \multicolumn{3}{l|}{AA4 (no MeerKAT), 133 antennas}\\
        Observation length & \multicolumn{3}{l|}{10 hours ($-5 < \mathrm{HA} < 5$)} \\
        Declination & -30$^\circ$ & & \\
        $uv$-range for analysis & \multicolumn{3}{l|}{$<10^5 \lambda$} \\
        Frequency band & 5b & 8.3 -- 10.8 GHz & 12.9 -- 15.4 GHz \\
        Primary beam & FWHM (from Gaussian fit) & 7.25 arcmin & 4.89 arcmin \\
        Sensitivity & Naturally weighted image & 0.402 $\mu$Jy beam$^{-1}$ & 0.569 $\mu$Jy beam$^{-1}$\\
        Synthesized beam & Naturally weighted image & 3.3 $\times$ 3.0 arcsec & 3.2 $\times$ 2.8 arcsec\\
    \end{tabular}
    \caption{Instrumental parameters used for simulations.  The image sensitivities and synthesized beams quoted incorporate the given $uv$-range.}
    \label{tab:ska_params}
\end{table}

\subsection{Cluster model}\label{sec:model:prof}
As a baseline cluster model, we assume the \citet{Arnaud2010} (hereafter A10) model which relates cluster mass\footnote{As reference mass definition, $M_{500}$, we consider the mass enclosed within $r_{500}$, the radius within which the average density equals $500\times$ the cosmic critical density at the cluster's redshift.} $M_{500}$ and redshift to a generalized Navarro-Frenk-White (gNFW) pressure profile \citep{2007ApJ...668....1N}.  The pressure profile is given by

\begin{equation}
    P(r) = P_{500} \left [ \frac{M_{500}}{3 \times 10^{14} h_{70}^{-1} M_\odot} \right ]^{\alpha_P} \mathbb{P}(r),
\end{equation}

where $\alpha_P = 0.12$ is a fitted parameter encoding a variation of pressure profile with mass. $P_{500}$ is a characteristic pressure derived from self-similarity arguments \citep{2007ApJ...668....1N} and is given by

\begin{equation}
    P_{500}=1.65 \times 10^{-3} h(z)^{8/3} \left [ \frac{M_{500}}{3 \times 10^{14} h_{70}^{-1} M_\odot} \right ]^{2/3} h_{70}^2\, \mathrm{keV cm^{-3}.}
\end{equation}

$\mathbb{P}(r)$ is a normalized pressure profile given by

\begin{equation}
    \mathbb{P}(r) = \frac{P_0}{(r/r_s)^\gamma \left [ 1 + (r/r_s)^\alpha \right ]^{(\beta-\gamma)/\alpha}}
\end{equation}

where $r_s$ is a characteristic size for the cluster and is related to the physical cluster radius $r_{500}$ by another parameter $c_{500} \equiv r_{500}/r_s$.  A10 gives `universal' parameters $[P_0, c_{500}, \gamma, \alpha, \beta] = [8.403 h_{70}^{-3/2}, 1.177, 0.3081, 1.0510, 5.4905]$ based on their fit to a sample of nearby, relaxed clusters.  We use this model with these particular parameters unless stated otherwise.

\section{tSZ observations with the SKA}\label{sec:sim}

In this section, we present some simulations exploring situations where we expect that the high resolution and sensitivity of the SKA will provide constraints which are complementary to cluster observations with other instruments.

Due to the relatively small primary beam size of the SKA in Band 5b, we expect the primary use case will be follow up observations of clusters detected with other instruments rather than blind surveys.  Our simulations therefore assume prior knowledge of a cluster at the given position.

\subsection{Mass-redshift distribution}\label{sec:sim:massz}

\begin{figure}[!h]
    \centering
    \includegraphics[width=\linewidth]{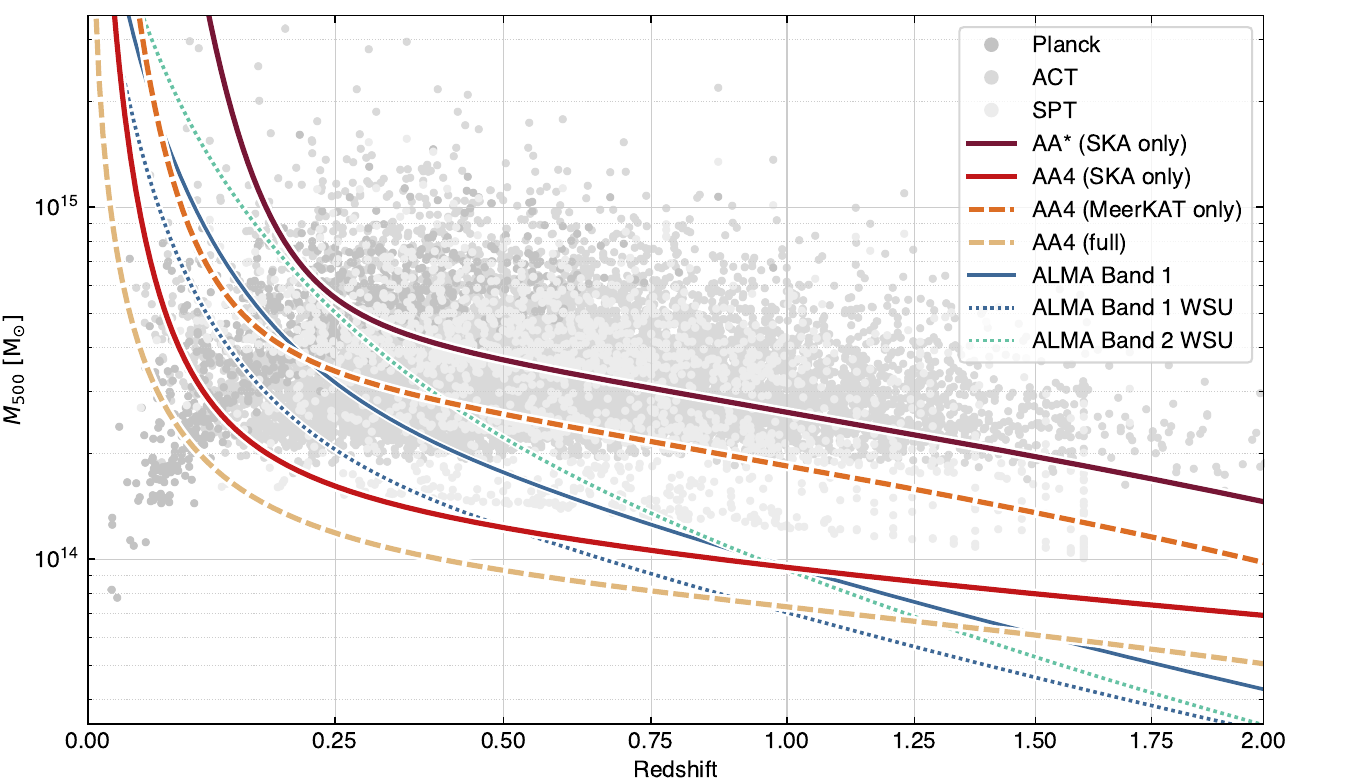}
    \caption{tSZ detection thresholds as a function of cluster mass and redshift for different SKA configurations, assuming a target significance of $5\sigma$, for 10 hours of on-source integration and a dual-band spectral setup ($2\times2.5~\mathrm{GHz}$; see Table~\ref{tab:ska_params} for details). The solid lines denote the sensitivity estimates for the nominal SKA-MID AA* and AA4 array configuration, i.e., comprising only the 15m-dish subarray (``SKA only''). To assess prospects for an enhancement of SKA-MID capabilities for SZ applications, we further plot as dashed lines the sensitivity limits obtained when considering Band 5b receivers on the MeerKAT 13.5m antennae as a distinct subarray (``MeerKAT only'') and correlated with the 15m dishes (``full''). For comparison, we further show as gray points the galaxy clusters from the available SZ wide-field surveys: \textit{Planck} \citep{Planck2016}, SPT/SPT-3G \citep{Bleem2020,Bleem2023,Kornoelje2025}, ACT \citep{Hilton2021,ACT2025}. For consistency with the observing constraints for SKA, we included only the galaxy clusters with declination $<30^{\circ}$. As additional reference, we plot the detection threshold curves for ALMA with the current Band 1 configuration, as well as with forecasts for Band 1\&2 after the full deployment of the WSU \citep{2023pcsf.conf..304C}. All the ALMA simulations are computed assuming the most compact configuration available for the 12-meter array.}
    \label{fig:masszeta}
\end{figure}

As broadly discussed in the introduction, among the key advantages of the SZ effect is the redshift independence of its surface brightness. This opens the possibility of probing the structure and thermodynamics of the ICM from the local Universe to the most massive systems at $z\simeq2$ in a nearly uniform way across cosmic time. In the context of radio-interferometric observations of the tSZ effect, the main limitations to the detection of the tSZ effect are posed by the combination of the finite instrumental sensitivity and the mismatch between the range of angular scales probed in a given observations and the one characteristic of the targeted clusters. To forecast the capabilities of SKA-MID in mapping the tSZ signature of galaxy clusters, fully accounting for its transfer function and expected noise properties, we derive mass-redshift detection thresholds for different SKA-MID configurations (Fig.~\ref{fig:masszeta}). As reference observing setup, we consider an on-source integration time of 10 hours and two spectral bands each covering a $2.5~\mathrm{GHz}$ bandwidth (consistent with Table~\ref{tab:ska_params}). 

To estimate the expected significance level, we bootstrap over realizations of a $\chi^2$ minimization operation assuming a matched filter detection experiment. For a galaxy cluster with given mass $M_{500}$ and redshift $z$, we generate 1,000 model visibilities by projecting the SZ signal obtained from the corresponding self-similar A10 universal pressure profile (see Sect.~\ref{sec:model:prof}) and injecting independent noise realizations. For each resulting mock observation, we compute the expected signal-to-noise ratio under the assumption of using the same A10 tSZ template as filter kernel, reducing the detection problem to a one-parameter fit for the overall tSZ amplitude. We then average over the resulting distribution of signal-to-noise estimates to obtain the reference integrated significance for the given model. To map the range of cluster masses and redshift, we repeat this operation over a fine grid of ($M_{500}$,$z$) values. In Fig.~\ref{fig:masszeta} we show as solid lines the resulting $5\sigma$ thresholds for the AA* and AA4 configurations. Although the AA* setup will allow for probing a large fraction of the massive galaxy clusters in the redshift range $z\simeq0.25-2$, the extension to AA4 will open the possibility of a systematic tSZ exploration of the ICM properties of galaxy clusters over the entire mass-redshift regime probed by wide-field survey experiments. 

In addition to the planned AA* and AA4 setup, we include forecasts for the SKA-MID AA4 capabilities when integrating the Band 5b receivers on the MeerKAT antennae, operating either as a standalone subarray or in correlation with the SKA dishes.  The inclusion of MeerKAT antennae will be discussed further in Section~\ref{sec:discussion:meerkat}.

As the main point of comparison, we perform a similar analysis to derive detection thresholds for ALMA given the same 10-hour observation length. Currently, ALMA is the facility that is most similar to SKA in terms of observing capabilities (i.e., achievable angular resolution, optimization for targeted SZ observations instead of wide-field detection surveys). For our simulations, we specifically consider observations in Band 1, currently providing the highest SZ sensitivity among the available spectral bands. Due to a combination of sensitivity, amplitude of the tSZ signal, and range of spatial scales probed in an observation, other bands and other configurations would result in mass thresholds higher than reported for Band 1. We further show the expected capabilities after the integration of the Wideband Sensitivity Upgrade \citep[WSU;][]{2023pcsf.conf..304C}, for Band 1 as well as for Band 2, the only WSU-ready spectral band to date. For all the reported cases, we simulate the observations adopting the most compact configuration for the main 12-meter array. Further, we do not include any complementary measurements with the compact 7-meter array, as we consider the case where large-scale information could be integrated from wide-field survey data. Although AA* and the MeerKAT subarray will be limited to higher mass than what the current ALMA capabilities can reach, the comparison clearly shows that the SKA-MID AA4 will be equally or more competitive than ALMA in terms of SZ detection sensitivity for low- to intermediate-redshift clusters ($z\lesssim1$). This poses SKA-MID as a key facility for future SZ studies, capable of extending targeted cluster observations to low-mass clusters and providing a natural counterpart to state-of-the-art high-resolution SZ facilities.

We would like to provide some final remarks. First, despite the competitive mapping speed in comparison to other high-frequency tSZ instruments, SKA-MID will still be limited to targeted observations, complementing the capability of state-of-the-art and prospective wide-field survey facilities. Second, the simulations are performed assuming no contamination from synchrotron emitting sources. This will be discussed further in Section~\ref{sec:discussion:synch}.

Finally, in our analysis, we are assuming the pressure distribution of the simulated clusters to be spherical and to follow an A10 radial profile. Clearly, any deviations from spherical symmetry (as in the case of merging systems) could bias the detectability of the SZ signal. At the same time, departures from the universal profile can be expected either due to the specific dynamical state of the system or to general evolution of the average pressure distribution with respect to the self-similar expectation. The latter is particularly relevant in the low-mass end of the cluster/group population, where feedback processes can dominate the thermal evolution of the ICM haloes. As we will discuss in Sect.~\ref{sec:sim:profiles}, though, the tSZ sensitivity and resolving power of SKA will provide the means for quantifying the level of these deviations and building a comprehensive model of ICM pressure.

\subsection{Pressure profile constraints}\label{sec:sim:profiles}

Under a simple gravitational collapse model, cluster pressure profiles are expected to be self-similar (e.g.\ \citealt{1985ApJS...58...39B}, \citealt{1986MNRAS.222..323K}).  Understanding how observed pressure profiles deviate from this expectation as a function of mass, redshift and/or merger history helps to shed light on the cluster formation process and how baryonic physics alters the simple gravitational collapse process (e.g.\ \citealt{2000MNRAS.317.1029P}, \citealt{2012MNRAS.421.1583M}, \citealt{2023MNRAS.523.1228L}).  Typically, pressure profiles are fit by averaging over a sample of clusters (assuming, in general, spherical symmetry), using tSZ data for the outskirts and X-ray pseudo-pressure for the inner parts where high resolution is needed (e.g. \citealt{2019A&A...621A..41G}).  However, it is known that X-ray constraints are subject to biases from clumping, and lack of precision in temperature due to the requirement to fit a spectrum to extract this information.  Pressure profile fits using tSZ data alone are therefore desirable.

As a case study, we simulate clusters with the pressure profiles derived in \citet{2014ApJ...794...67M} using a sample of high-redshift SZ-selected clusters observed in X-ray with \emph{Chandra}.  The properties of the average profiles for a relaxed and un-relaxed subsample are summarized in Table~\ref{tab:mcdonald14}.

\begin{table}
    \centering
    \begin{tabular}{lccccccc}

\hline\hline
Subsample & $\left< z \right>$ & $\left<M_{500} \right>$ & $P_0$ & $c_{500}$ & $\gamma$ & $\alpha$ & $\beta$ \\\hline
High-z, CC & 0.82 & 4.2 & 3.70 & 2.80 & 0.21 & 2.30 & 3.34 \\	
High-z, NCC & 0.82 & 4.2 & 3.91 & 1.50 & 0.05 & 1.70 & 5.74 \\	
\hline
    \end{tabular}
    \caption{Parameters used for cluster pressure profile constraint simulations, taken from \citet{2014ApJ...794...67M}.  The two subsamples are cool-core (`CC') and non-cool-core (`NCC').  In the real cluster samples, the mean mass and redshifts were slightly different between the subsamples; here we keep them the same as the overall means for simplicity.}
    \label{tab:mcdonald14}
\end{table}

We analyze the cluster simulations using \textsc{McAdam} \citep{2009MNRAS.398.2049F}, an analysis software designed specifically for interferometric observations of the tSZ effect.  \textsc{McAdam} employs a forward-modelling approach to perform Bayesian analysis and produce cluster model parameter constraints, calculating the likelihood in the $uv$-plane to exploit the uncorrelated, Gaussian-distributed nature of the noise.

We consider two scenarios for deciding which parameters to vary in our analysis of the simulated clusters.  In the first scenario, the cluster has been detected and characterized by a lower-resolution SZ survey, so we have constraints on the integrated Compton-$y$ parameter and on $\beta$, which defines the slope of the pressure profile at large radius.  In the second scenario, the cluster has been detected in some other waveband and so we wish to assume minimal prior information on any of the pressure profile parameters.  We give the priors that we assume in these two scenarios in Table~\ref{tab:pressure_priors}.

\begin{table}
    \centering
    \begin{tabular}{lcc}
    \hline\hline
Parameter & Prior (scenario 1) & Prior (scenario 2) \\\hline
$\theta_\mathrm{s}$ / arcmin & $U[0.0025, 5]$ & $U[0.0025, 5]$ \\
$Y_\mathrm{tot}$ / arcmin$^2$ & $\mathcal{N}(\hat{Y_\mathrm{tot}}, 0.2 \hat{Y_\mathrm{tot}})$ & $U[10^{-5}, 10^{-3}]$ \\
$\gamma$ & $U[0.0, 1.0]$ & $U[0.0, 1.0]$ \\
$\alpha$ & $U[0.1, 3.5]$ & $U[0.1, 3.5]$ \\
$\beta$ & $\delta(\hat{\beta})$ & $U[3.0, 7.5]$ \\\hline
    \end{tabular}
    \caption{Priors used on parameters for our pressure profile analysis, in the two scenarios outlined in the text.  We use an observational formulation of the GNFW model for this analysis; $\theta_\mathrm{s}$ is the angular equivalent to $r_\mathrm{s}$, and $Y_\mathrm{tot}$ represents the total Compton-$y$ parameter integrated out to infinity.  $U[x_1,x_2]$ indicates a uniform prior between limits $x_1$ and $x_2$, while $\mathcal{N}(\mu,\sigma)$ represents a normal distribution.  $\hat{x}$ is used to indicate the true (input) value of a parameter used to create the simulation.}
    \label{tab:pressure_priors}
\end{table}

\begin{figure}
    \centering
    \includegraphics[width=\textwidth]{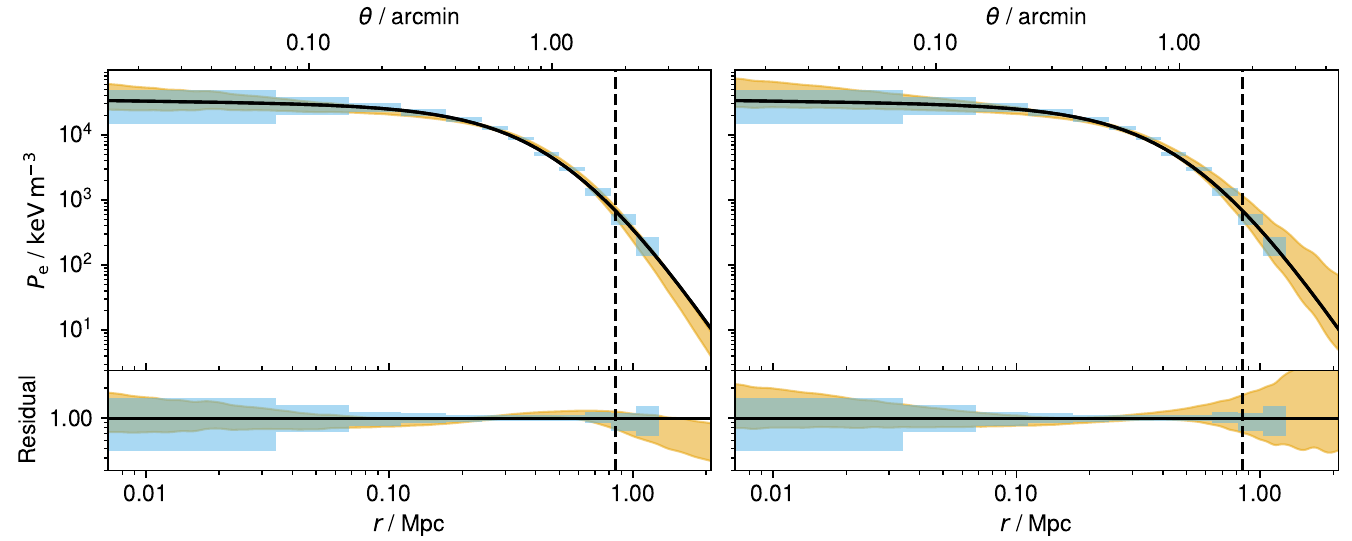}
    \caption{Pressure profile constraints derived from a simulated 10 hour observation of a non-cool-core cluster with the SKA (orange bands) compared to the constraints from the stacked X-ray observations of the 20-cluster non-cool-core sample from \citet{2014ApJ...794...67M} (blue bins).  Both SKA 5b subbands are included in these simulations.  In the left-hand plot, prior constraints on the large-scale cluster parameters have been assumed; in the right-hand plot, all parameters are varied.  The black vertical dashed line shows the cluster $r_{500}$.}
    \label{fig:pressure_profiles_1day}
\end{figure}

\begin{figure}
    \centering
    \includegraphics[width=\textwidth]{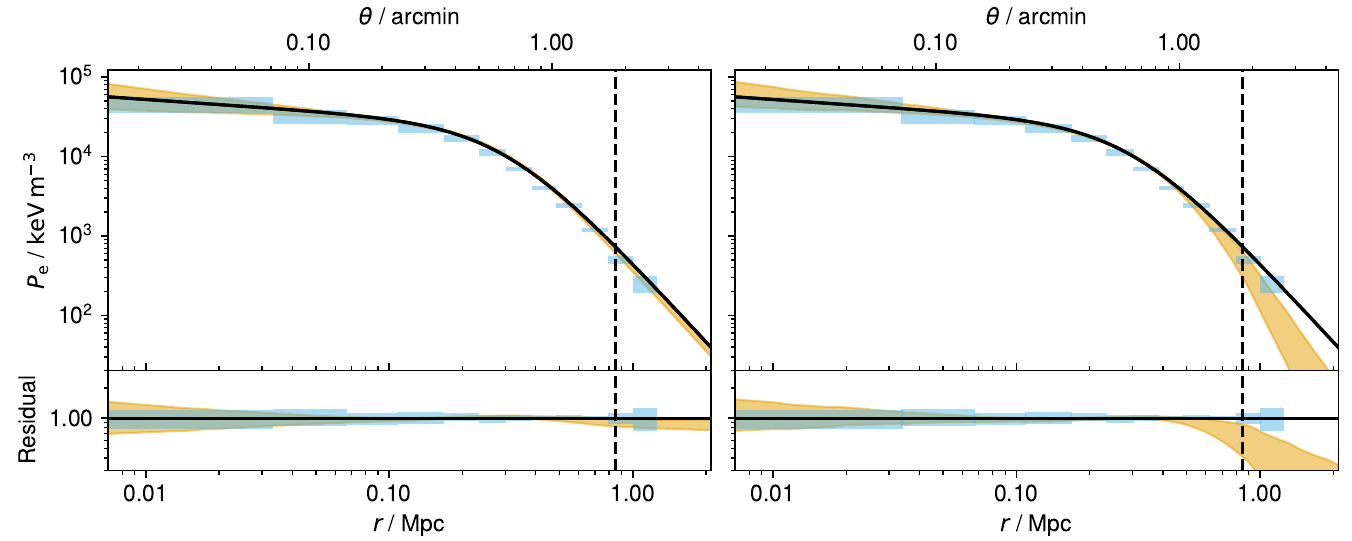}
    \caption{Pressure profile constraints derived from a joint analysis of simulated 10$\times$ 10 hour observations of a cool-core cluster with the SKA (orange bands) compared to the constraints from the stacked X-ray observations of the 20-cluster cool-core sample from \citet{2014ApJ...794...67M} (blue bins).  Only the upper SKA subband is included in these simulations.  In the left-hand plot, prior constraints on the large-scale cluster parameters have been assumed; in the right-hand plot, all parameters are varied.  The black vertical dashed line shows the cluster $R_{500}$.}
    \label{fig:pressure_profiles_10days}
\end{figure}

Results are presented in Figures~\ref{fig:pressure_profiles_1day} and \ref{fig:pressure_profiles_10days}.  Figure~\ref{fig:pressure_profiles_1day} shows the inferred constraints on the pressure profile of a non-cool-core cluster based on a single 10-day observation, using both 5b subbands, under the two scenarios.  For comparison, we also show the pressure profile constraints based on the 20-cluster sample from \citet{2014ApJ...794...67M}.  The SKA observation of a single cluster gives constraints that are comparable with the constraints from the stacked X-ray observations, particularly in the core where the X-ray constraints suffer from the lack of precision in the temperature measurements.  The constraints are clearly better in the outskirts when prior information on the large-scale properties is incorporated, however even in scenario 2 with the uninformative priors useful information is recovered out to $\sim r_{500}$.  We find that the inclusion of the lower-frequency subband aids slightly in the recovery of the larger scales, however most of the pressure profile is reconstructed well using the higher-frequency subband only.  This means that the lower-frequency subband can be leveraged in the case that diffuse non-thermal emission must be disentangled from the tSZ effect; this will be discussed further in Section~\ref{sec:discussion:synch}.

Figure~\ref{fig:pressure_profiles_10days} shows the results from fitting 10 realizations of the cool-core cluster simulation simultaneously, emulating a stacked fit to a sample of 10 clusters, or alternatively a 100-hour observation of a single cluster.  In this case we have used only the higher-frequency subband.  The pressure profile constraints derived are comparable to or better than the 20-cluster sample X-ray constraints (which are much more precise in the core than the corresponding non-cool-core X-ray constraints due to the strong X-ray emission from the cool, dense core) over all of the radial range.  In scenario two, the large-scale constraints are biased downward.  This is due to the combination of degeneracies inherent in the GNFW model (see \citet{2019MNRAS.486.2116P} for more detail on these issues) and could be alleviated by placing a more restrictive prior on $Y_\mathrm{tot}$ based on the cluster's redshift and estimated mass, and/or using a non-parametric model (e.g. \citealt{2018MNRAS.481.3853O}).

We note that these are idealized simulations not including effects such as departures from spherical symmetry, imperfect removal of radio sources, etc.  These constraints therefore represent a best-case scenario, but give promising indications for the potential of the SKA in constraining high-redshift cluster pressure profiles.

\subsection{Turbulent pressure fluctuations}\label{sec:sim:perturb}

Galaxy clusters grow through subcluster mergers and continuous accretion from the cosmic web, which drive shocks that heat the intracluster gas. In cluster cores, feedback from active galactic nuclei (AGN) further injects energy via episodic outbursts. A fraction of the energy released, whether from large-scale structure formation or from AGN activity, cascades into turbulence that dissipates and contributes to gas heating. Turbulence is expected to be the dominant source of non-thermal pressure support in the ICM \citep{Vazza2016}, and thus a major contributor to the hydrostatic mass bias affecting mass estimates derived from X-ray and SZ observations under the assumption of thermal hydrostatic equilibrium \citep{Angelinelli2020}. This bias remains one of the main challenges for precision cluster cosmology. 

Turbulence in the ICM can be directly measured through high-resolution X-ray spectroscopy by detecting the Doppler shift and broadening of emission lines. This has recently become possible with the Hitomi and XRISM satellites. However, such observations are extremely resource-intensive and therefore limited to the cores of massive nearby clusters \citep{Hitomi2016,XRISM2025a,XRISM2025b}. As an alternative, turbulence can be indirectly probed by analyzing thermodynamic fluctuations driven by the turbulent velocity field \citep{Gaspari2014,Zhuravleva2023}. This approach has been applied in X-rays, which trace ICM density fluctuations \citep{Dupourque2024,Heinrich2024}, and more recently through the tSZ effect, which is sensitive to pressure fluctuations \citep{Khatri2016,Romero2023,Adam2025}.

In this section, we assess the sensitivity of the SKA to the power spectrum of pressure fluctuations in galaxy clusters, as a tracer of turbulence. A comprehensive sensitivity analysis across cluster mass, redshift, and input power spectrum parameters is beyond the scope of this study. Instead, we focus on evaluating how well SKA can recover a known input power spectrum using a representative test case. Specifically, we simulate the signal from a massive cluster with $M_{500} = 7 \times 10^{14}$ M$_{\odot}$ at redshift $z = 0.5$, a configuration chosen to match the spatial scales accessible to SKA tSZ observations.

We model the ICM pressure as the sum of a smooth thermal hydrostatic component and pressure fluctuations,
\begin{equation}
    P(x_1, x_2, x_3) = \bar{P}(r) + \delta P(x_1, x_2, x_3),
    \label{eq:p_turb_modeling}
\end{equation}
where $\bar{P}(r)$ denotes the radial pressure profile and $\delta P(x_1, x_2, x_3)$ the pressure fluctuations. Here, $x_{1,2,3}$ refers to the spatial coordinates and $r$ to the radius in 3D. The cluster mean pressure profile is modeled using the generalized NFW (gNFW) profile corresponding to the morphologically disturbed case from \citet{Arnaud2010}. The pressure fluctuations are modeled as lognormal variations following the power spectrum defined by
\begin{equation}
    \mathcal{P}_{\delta P/\overline{P}}(k_{\rm 3D})=
    \sigma_\mathcal{P}^2\frac{k_{\rm 3D}^\alpha {\rm exp}\left(-\frac{1}{k_{\rm 3D}^2L_{\rm inj}^2}\right){\rm exp}\left(-k_{\rm 3D}^2L_{\rm dis}^2\right)}{\int 4 \pi k_{\rm 3D}^{\alpha+2} {\rm exp}\left(-\frac{1}{k_{\rm 3D}^2L_{\rm inj}^2}\right){\rm exp}\left(-k_{\rm 3D}^2L_{\rm dis}^2\right){\rm d}k_{\rm 3D}}.
    \label{eq:P_turb_pk3d}
\end{equation}
The parameter $\sigma_\mathcal{P}$ represents the root-mean-square amplitude of the pressure fluctuations, $\alpha$ the spectral slope in the inertial range, $L_{\rm inj}$ the injection scale, and $L_{\rm dis}$ the dissipation scale. We adopt $\sigma_\mathcal{P} = 0.4$ and $L_{\rm inj} = 0.4\,R_{500}$, following ensemble-average constraints derived from the NIKA2 LPSZ sample \textcolor{black}{(Adam et al., in prep.)}. These values correspond to a kinetic-to-thermal pressure ratio of approximately 14\%, based on the scaling relation from \citet{Zhuravleva2023}, and thus a similar value of the hydrostatic mass bias. The spectral slope is fixed to $\alpha = -11/3$, consistent with a Kolmogorov cascade \citep{Kolmogorov1941}, and the dissipation scale is set to $L_{\rm dis} = 1~\mathrm{kpc}$. The latter choice has negligible impact on the results, as it lies well below the observational sensitivity limits.

\begin{figure}[!ht]
    \centering
    \includegraphics[width=1\textwidth]{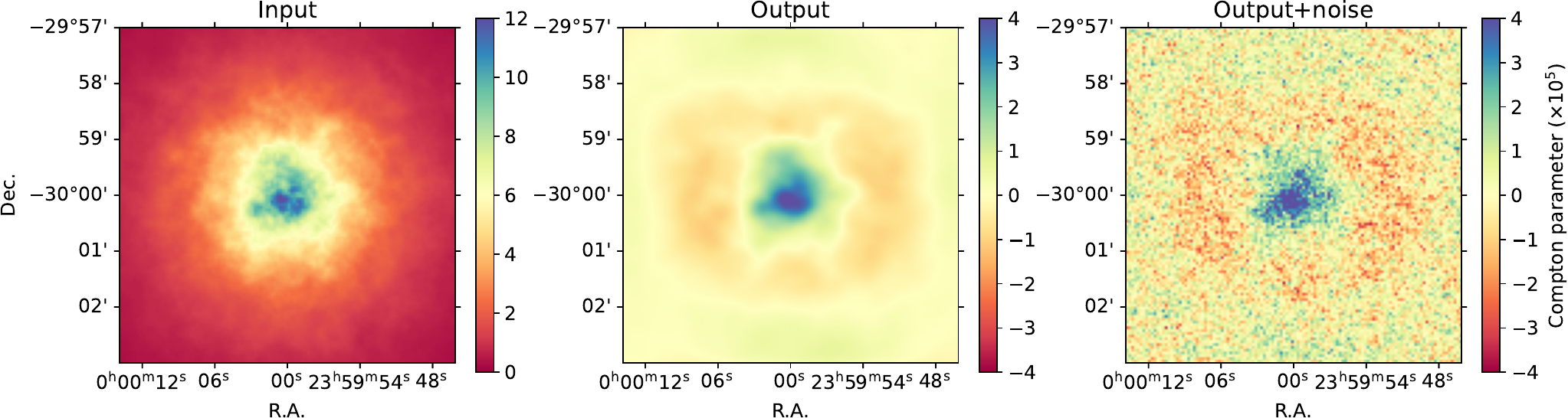}
    \caption{Mock cluster simulation in the presence of SZ fluctuations. Left: input SZ Compton parameter map. Middle: output map without noise contribution. Right: output map with noise contribution.}
    \label{fig:mock_cluster_turbulence}
\end{figure}

The mock Compton-$y$ images of the clusters are generated using the \texttt{PITSZI} software \citep{Adam2025}, through the \texttt{Model} class, and subsequently processed with the SKA mock observation pipeline using the parameters listed in Table~\ref{tab:ska_params}. Figure~\ref{fig:mock_cluster_turbulence} illustrates an example of the input mock cluster and the corresponding reconstructed images obtained after simulated observations, both with and without instrumental noise. We note that the input signal is dominated by the smooth radial pressure profile, while secondary tSZ fluctuations trace the underlying pressure inhomogeneities. The reconstructed map is affected by the SKA instrument response, which suppresses large-scale modes and smooths small-scale structures, producing characteristic ringing around the cluster. Under the adopted observational setup, the cluster is well recovered even in the presence of noise, and deviations from spherical symmetry are visible by eye.

\begin{figure}[!ht]
    \centering
    \includegraphics[width=\textwidth]{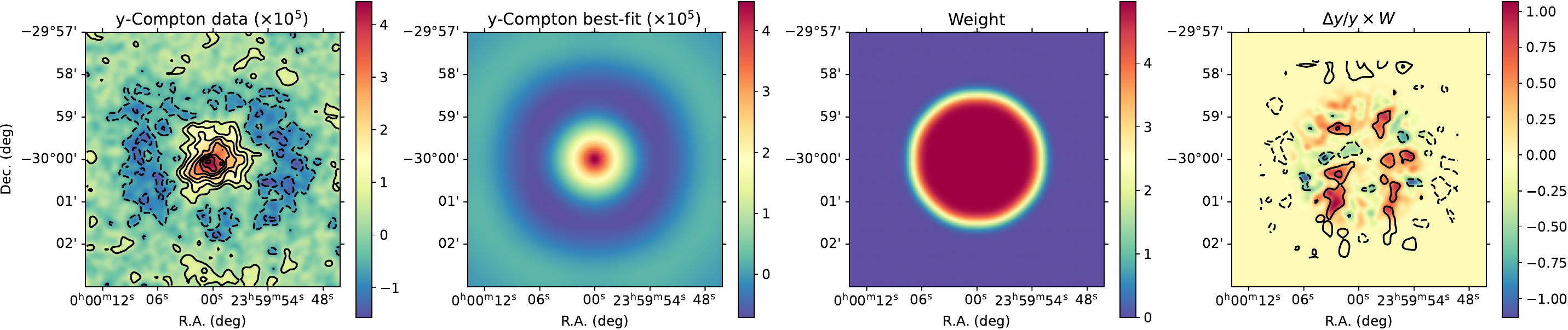}
    \caption{tSZ fluctuation data. 
    Left: SKA mock Compton parameter map. The map was smoothed with a 10 arcsec Gaussian (FWHM) for visual purpose and contours are every $2 \sigma$.
    Middle left: Compton parameter model for the thermal hydrostatic component.
    Middle right: weight map $W$.
    Right: tSZ fluctuation map, multiplied by the weight.}
    \label{fig:input_data_turbulence}
\end{figure}

We analyze the mock data in real space using the \texttt{InferenceFluctuation} subpackage of \texttt{PITSZI}, which performs forward fitting of the relevant fluctuation parameters. The tSZ fluctuation map, $\Delta y / y \times W$, is computed as the difference between the Compton-$y$ map and the thermal hydrostatic model, normalized by the latter. The region of interest is defined as a circular aperture of radius $R_{500}/2$, where the sensitivity to fluctuations is optimal; this region is used to construct the weight map $W$. The Compton-$y$ map, thermal hydrostatic model, weight map, and resulting tSZ fluctuation map are shown in Figure~\ref{fig:input_data_turbulence}.
The power spectrum of the tSZ fluctuations is extracted and compared to a pressure fluctuation model that incorporates line-of-sight projection, instrumental effects, and observational noise. The filtering introduced by the instrument is quantified through simulations and applied as an effective transfer function. The tSZ fluctuation power spectrum is computed over the range $k_{\rm 2D} = 1/\theta_{500} \simeq 1/3~\mathrm{arcmin}^{-1}$ to $k_{\rm 2D} = 1/10~\mathrm{arcsec}^{-1}$, linearly binned into 30 intervals. The noise contribution is estimated via Monte Carlo simulations, and the full noise covariance matrix is included, accounting also for the variance of the input model due to the stochastic nature of the signal. The resulting power spectrum model is then given by
\begin{equation}
    M(k_{\rm 2D}) = \mathcal{P}_{\delta y/\overline{y}}(\sigma_\mathcal{P}, L_{\rm inj}) + A_{\rm noise} \times \mathcal{P}_{\rm noise},
    \label{eq:p_turb_modeling2}
\end{equation}
where we fit for the pressure fluctuation power spectrum parameters, $\sigma_\mathcal{P}$ and $L_{\rm inj}$, as well as $A_{\rm noise}$, the normalization of the noise contribution over which we marginalize.

\begin{figure}[!ht]
    \centering
    \includegraphics[height=6cm]{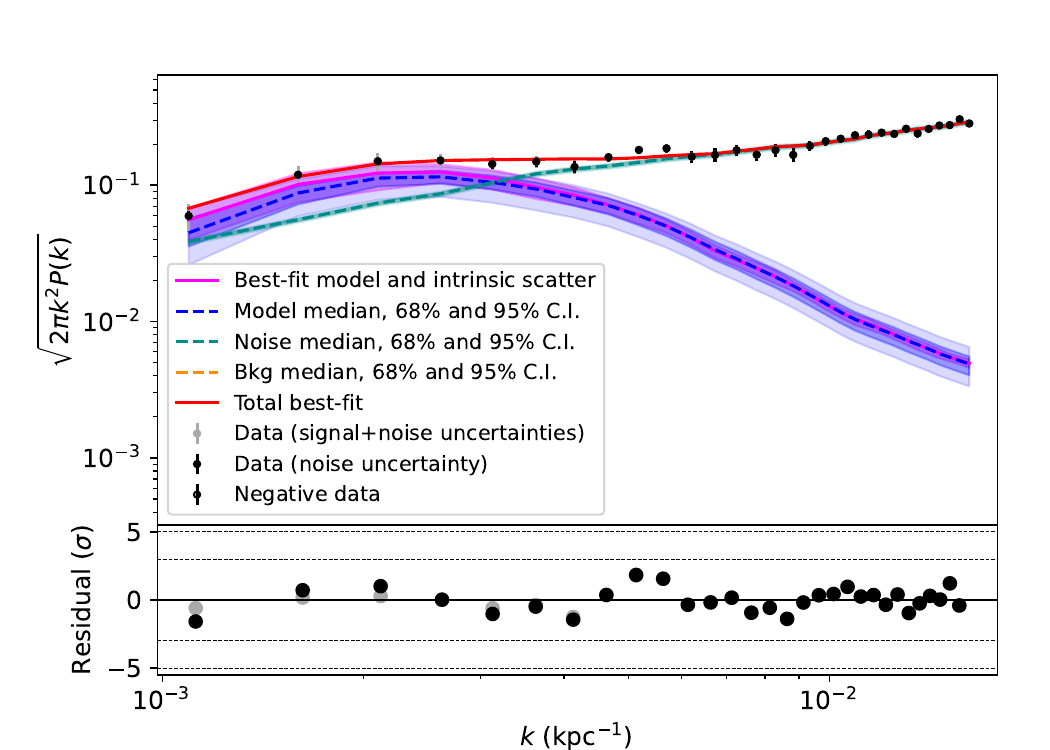}
    \includegraphics[height=6cm]{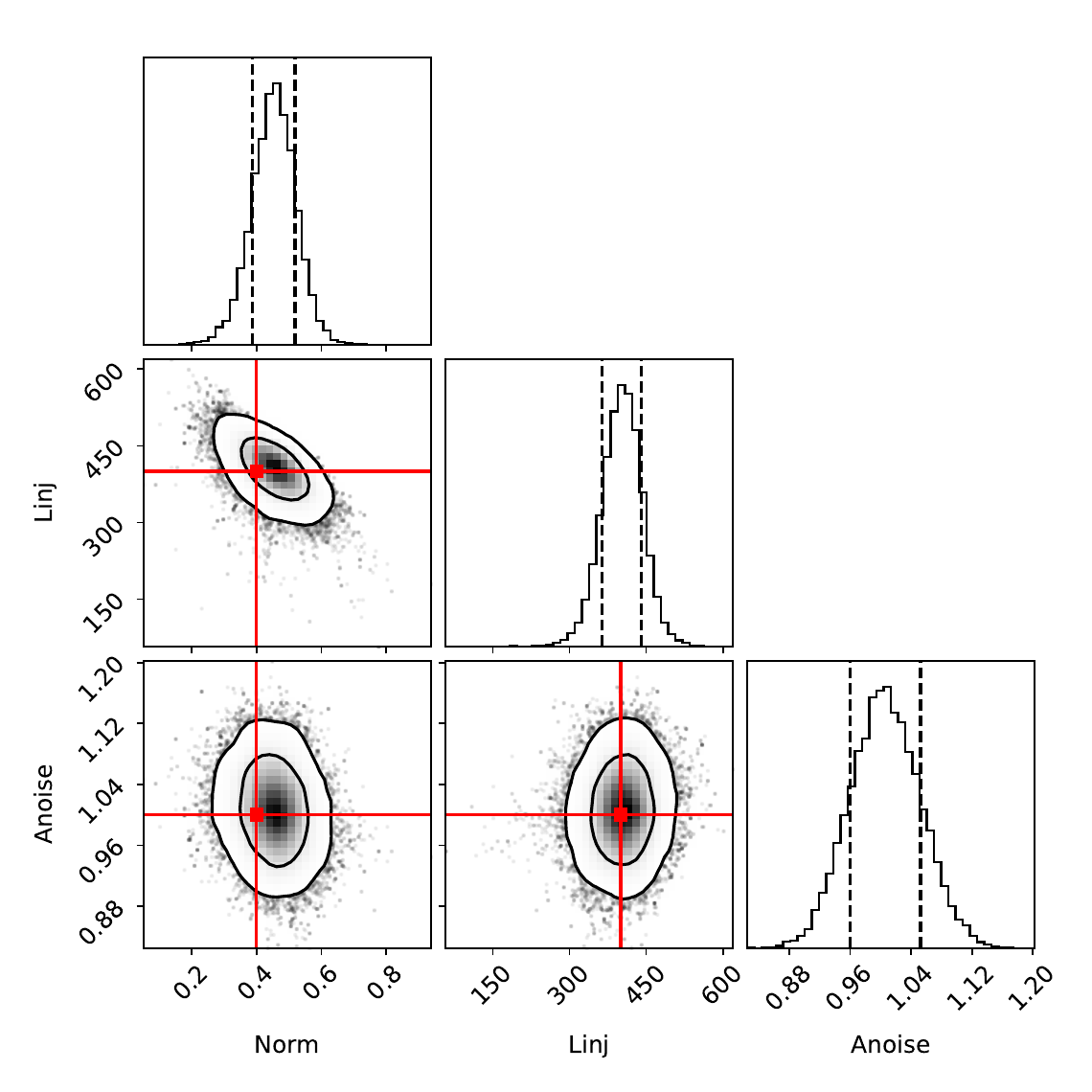}
    \caption{Left: 2D power spectrum of the tSZ fluctuations. The data points are given in black. We report the error bars arising from the noise only (black) and when accounting for the model variance (grey). The green line and contours correspond to the noise $A_{\rm noise} \times \mathcal{P}_{\rm noise}$. The blue line and contours are for the tSZ fluctuation median confidence interval ($\mathcal{P}_{\delta y/\overline{y}}(\sigma_\mathcal{P}, L_{\rm inj})$). The magenta line and shaded region give the model best-fit and intrinsic scatter (model intrinsic variance only), which increases on large scales. The red line gives the total best-fit model $M(k_{\rm 2D})$. In the given setup, no extra background was included and the data are all positive because we are not considering the cross-spectra of two datasets.
    Right: recovered constraint in the parameter space. Contours correspond to 68\% and 95\% confidence. The red crosses give the input parameter values.
    }
    \label{fig:output_cluster_turbulence}
\end{figure}

In Figure~\ref{fig:output_cluster_turbulence} (left panel), we show the measured tSZ fluctuation power spectrum along with the noise contribution. The right panel presents the corresponding constraints in the parameter space. Under the adopted observational setup, SKA exhibits excellent sensitivity to the pressure fluctuation power spectrum, even for a single cluster and accounting for intrinsic model variance. For this input model, the normalization is recovered to within $\sim15\%$, and the injection scale to within $\sim10\%$.

Although this study is limited to a single test-case cluster, it demonstrates that SKA will be highly sensitive to pressure fluctuations in the ICM with just a few hours of observations.

\subsection{Merger-driven shocks}\label{sec:sim:shock}

Shock fronts represent the large-scale counterpart to tSZ fluctuations induced by merger activity throughout cluster evolution (see also Sect.~\ref{sec:sim:perturb} above) and have a profound impact on evolution of ICM halos \citep{Markevitch2007}. Their propagation initiates the thermalization of the kinetic energy injected in the ICM by merger episodes, while inducing local deviations in the overall equilibrium of ICM particle populations (in the form of, e.g., different ionization and temperature states for the ions and electrons in the post-shock region; \citealt{Vink2015}). Concurrently with turbulence, shock ICM compression plays a central role in the process of (re)acceleration of the non-thermal population of ICM electrons up to relativistic energies \citep{Brunetti2014}. While widely expected, though, mergers with a clean geometry and clearly defined shock fronts (as opposed to shock-heated gas) are rare. This is due to a combination of projection and geometrical effects, as well as of the specific dynamical configuration of a merger event. As a result, shock fronts have been unambiguously detected in only a handful of galaxy clusters \citep{Markevitch2006,Bourdin2013,Wang2016,Russell2022,Norseth2025} and mostly in the X-ray band, leaving the specific mechanisms controlling the shock-driven heating largely unconstrained.

\begin{figure}[!h]
    \includegraphics[width=\linewidth]{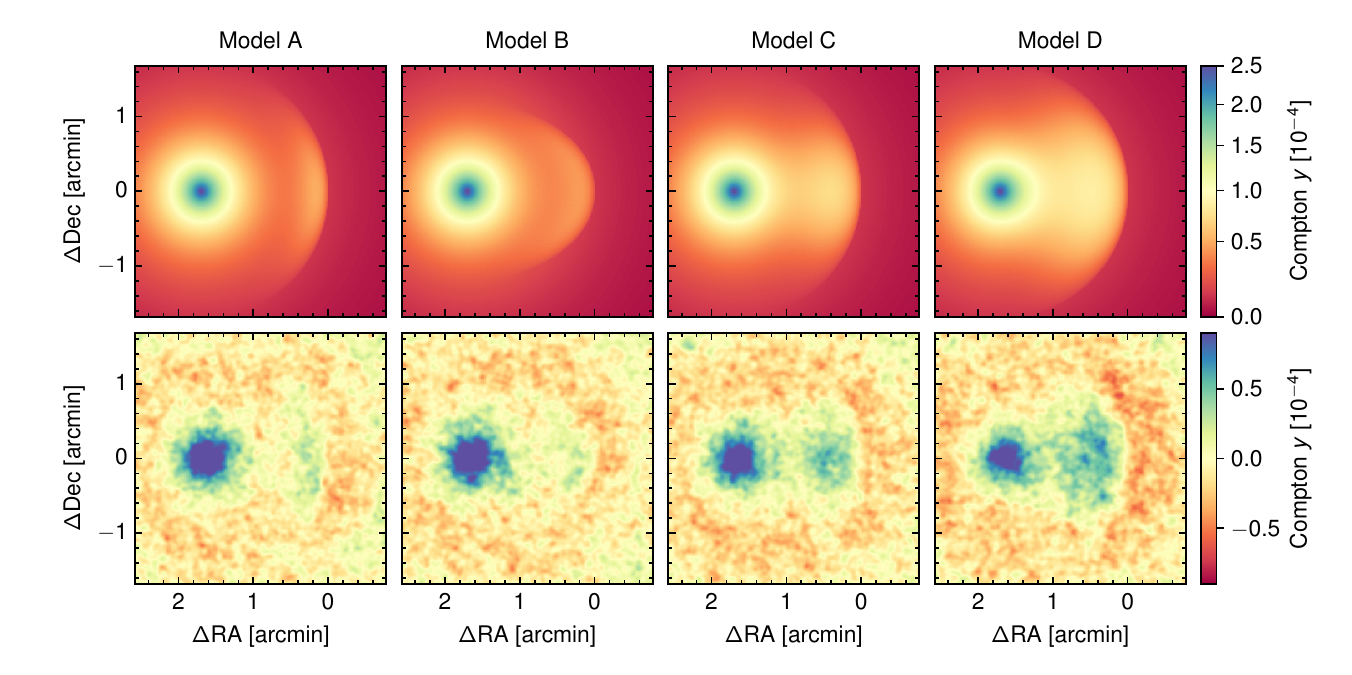}
    \caption{Selection of shock models employed in the forecast analysis. For all panels, we are considering a cluster with mass $M_{500}=5\times10^{14}~\mathrm{M_{\odot}}$ at redshift $z=1$, and a shock front with a Mach number $\mathcal{M}=3$ at a distance $r_{500}$ from the cluster centre. Model A: spherical shock cone, square cosine azimuthal variation of the Mach number. Model B: same as A, but for an ellipsoidal shock surface. Model C: same as A, but for a shallower post-shock pressure profile. Model D: same as C, but for a shallower azimuthal variation of the Mach number. The top row shows the input model while the bottom row shows a corresponding simulated SKA observation.  For visualization purposes, the images are generated adopting a $uv$ taper with an equivalent kernel width of $3''$.}
    \label{fig:shock_models}
\end{figure}

Here we explore the expected performance of SKA-MID in measuring features in the tSZ surface brightness of galaxy clusters associated with shock fronts and, in general, dynamically driven pressure discontinuities. To date, tSZ studies of shock fronts have been limited to a few, extreme individual cases \citep[e.g.][]{2013A&A...554A.140P,2015MNRAS.447.2497E,2016ApJ...829L..23B,2019A&A...628A.100D}, or to stacking experiments, \citep[e.g.][]{2024MNRAS.527.9378A}.  This is a direct consequence of the limited sensitivity or spatial dynamic range (or combination, thereof) of the available SZ facilities. The tSZ effect however offers key advantages compared to X-ray measurements. First, the linear dependence on ICM density allows for probing shock-driven discontinuities at large cluster-centric distances, providing a counterpart to the relic identification at radio wavelengths. Second, standard shock conditions predicts the density contrast between the upstream and downstream ICM to saturate asymptotically to a maximum factor $4$ increasing Mach number $\mathcal{M}$. On the other hand, the amplitude of shock-driven pressure jumps increases proportionally to $\mathcal{M}^2$, making the tSZ effect a much more sensitive probe of strong shocks compared to X-ray (see e.g. \citealt{2021A&A...651A..41C}). This may become particularly relevant when merger shocks are actually not homogeneous but show significant fluctuations on small spatial scales. X-ray surface brightness studies may reveal a fraction of the shock which only covers a small area of the shock surface but has high (spatially variable) Mach number. It should be noted, however, that radio emission related to acceleration of cosmic ray electrons may need to be disentangled. We will discuss this further in Sec.~\ref{sec:discussion:synch}. 

Still, the projected amplitude of pressure discontinuity depends severely on the specific merger configuration and morphological parameters (e.g., azimuthal geometry and strength variability, line-of-sight aperture radius of the shock surface, post-shock pressure evolution), and the same inference of the shock properties relies on our ability to disentangle their degenerate observational effects. We thus focus our analysis on constraining the expected significance level of shock-like structures in Band 5b AA4 observations by marginalizing over different shock configurations. In particular, we bootstrap over mock tSZ observations based on a simplified shock model. We consider the global ICM pressure distributions to be characterized by an A10 profile, and perturb the pressure structure by adding a three-dimensional shock front. For simplicity, we assume the shock surface to be rotationally symmetric with respect to an axis lying (i.e., assuming the merger to happen) on the plane of the sky. Further, we fix the Mach number and assume the pressure discontinuity to follow standard Rankine-Hugoniot jump conditions with instantaneous ion-electron re-equilibration. For each model iteration, we then randomly sample the following model parameters:
\begin{itemize}[leftmargin=*,parsep=0pt]
    \item \textbf{Cluster-centric distance.} In this experiment, we always consider a physical setup consistent with post-core passage scenario. We however change the distance of the shock front from the centroid of the global pressure distribution, reflecting the heterogeneity observed in real systems. 
    
    \item \textbf{Projected geometry.} Merger-driven shocks exhibit a great variety of geometries, as a result of the complex interplay of multi-scale plasma processes within the ICM, and of the non-linear assembly history of cluster haloes. To encode the impact of plane-of-sky geometry on the simulation, we parameterize the shock surface as an ellipsoidal section, and vary the ratio of the axes parallel and perpendicular to the axis of symmetry (see Models A and B in Fig.~\ref{fig:shock_models}).
    
    \item \textbf{Downstream pressure profile.} The injection of thermal energy in the post-shock region inherently makes the ICM pressure structure depart from a radial universal scaling. To avoid spurious discontinuities in the downstream region, we thus model the spatial evolution of the shock-driven pressure increase as $\delta P/P_{0}=\sin^{\alpha_1}(\frac{\pi}{2}x^{\alpha_{2}})\times h(x)$, where $x$ is the radial distance from the cluster centroid in units of shock cluster-centric distance $r_{s}$, and $h(x)$ is a top-hat function equal to one for $0<x<1$ and zero otherwise. The indices $\alpha_1$ and $\alpha_2$ control the steepness of the transition from the shock-enhanced to the universal ICM pressure (see Models A and C of Fig.~\ref{fig:shock_models}).
    
    \item \textbf{Azimuthal decrease of Mach number.} The properties of shocks increasingly deviate from an ideal planar case when considering off-axis regions of the shock front. The net effect is an azimuthal decrease of the amplitude of the shock-driven pressure discontinuity. Following, e.g., \citet{Wang2018}, we model the Mach number variation as a function of the azimuthal angle $\phi$ as $\cos^{\beta}(\phi/\phi_0)$, where $\phi_0$ and $\beta$ are free parameters of our model and control the azimuthal cut-off of the shock structure and steepness of the reduction (see Models C and B of Fig.~\ref{fig:shock_models}).
\end{itemize}

\begin{figure}[!h]
    \centering
    \includegraphics[clip,trim=0 10pt 0 0,width=\linewidth]{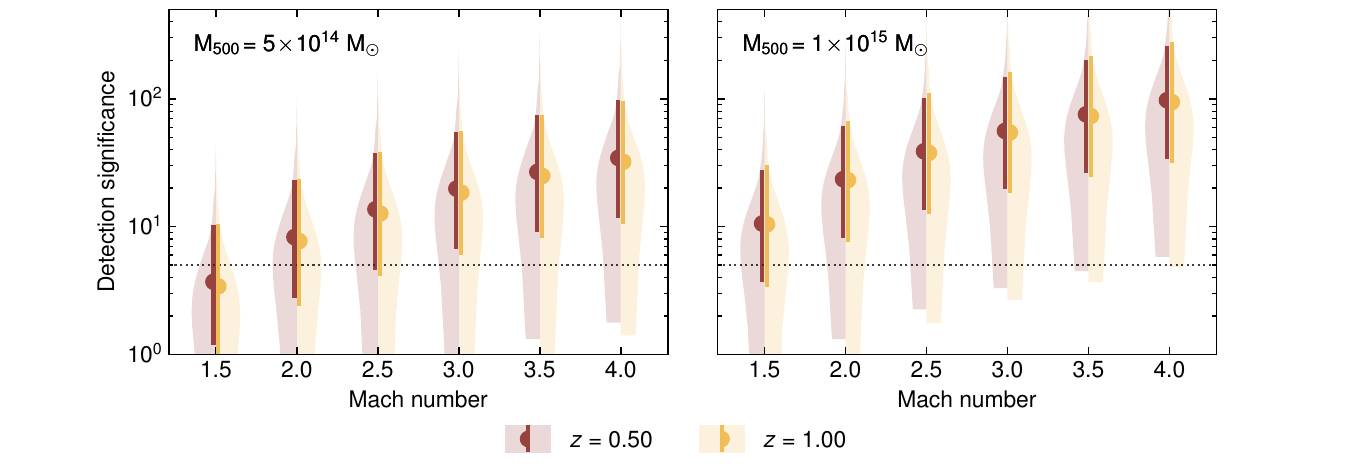}
    \caption{Detection significance of shock fronts as a function of Mach number for massive clusters ($M_{500}=5\times10^{14}~\mathrm{M_{\odot}}$, left panel; $M_{500}=1\times10^{15}~\mathrm{M_{\odot}}$, right panel) at redshift $z=0.50$ (red) and $z=1.00$ (yellow). The shaded regions summarize the distribution of significance level values obtained for the set of shock parameters explored in our analysis. The markers and the error bars denote the median and 16\textsuperscript{th} and 84\textsuperscript{th} percentile ranges of the distribution of significance estimates for each Mach number, respectively. The dotted horizontal line denotes the $5\sigma$ detection limit.}
    \label{fig:shocks}
\end{figure}
The result of this analysis are summarized in Fig.~\ref{fig:shocks}. To estimate the detection significance, we adopt the same optimization approach used in Sect.~\ref{sec:sim:massz}, but consider as matched kernel the projected tSZ model obtained from the shock-driven increase in the ICM pressure $\delta P_{\mathrm{e}}$ for each corresponding shock realization. Not surprisingly, the detection levels exhibit significant scatter for all the considered Mach numbers. At the same time, this analysis shows how SKA-MID will be capable of detecting the tSZ signal from shock fronts with a high signal-to-noise ratio ($\gtrsim10$) practically independently of redshift and for a relatively short integration time (10 hours on source; see Table~\ref{tab:ska_params} for further details).

\begin{figure}[!h]
    \centering
    \includegraphics[clip,trim=0 0 0 10pt,width=\linewidth]{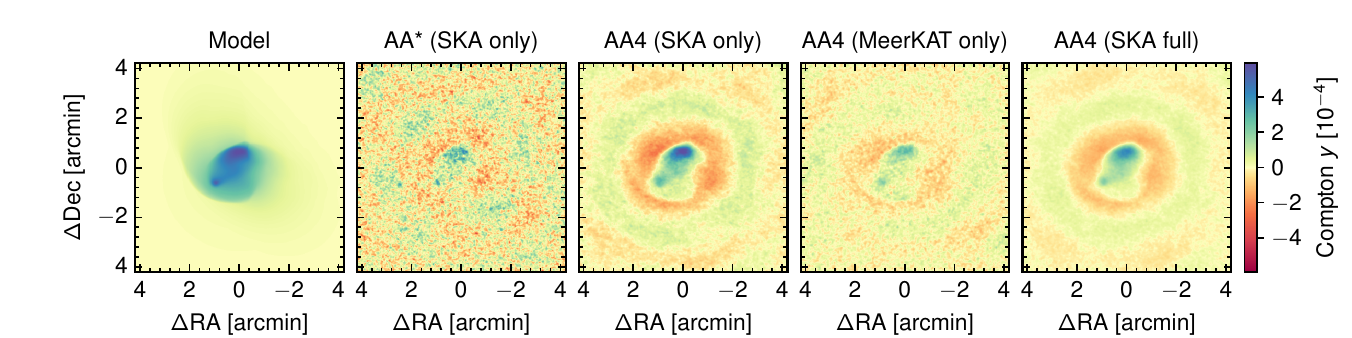}
    \caption{Mock observations of the SZ effect from El Gordo (ACT-CL J0102-4915, $z=0.870$; \citealt{Marriage2011}). The input model (first panel) is based on the hydrodynamical simulation by \citet{2015ApJ...813..129Z,2018ApJ...855...36Z}. The second and third panels show the mock images obtained from the nominal AA* and AA4 SKA-MID configurations. The fourth and fifth panels provide an additional comparison of the tSZ imaging capabilities of the MeerKAT antennae as a standalone subarray or correlated with the SKA dishes, respectively.}
    \label{fig:elgordo}
\end{figure}

Clearly, the characterization of ICM shock fronts will not be limited to their identification. The extensive spatial coverage will allow for the detailed tSZ mapping of ICM structures and spatially resolved inference of their thermodynamic properties. To derive realistic forecasts, we move beyond the simplified model employed above and consider a simulated equivalent of a known merging system. In particular, we consider the hydrodynamical simulation by \citet{2015ApJ...813..129Z,2018ApJ...855...36Z}, specifically tuned to reproduce the observational properties of the famous galaxy cluster El Gordo (ACT-CL J0102-4915, $z=0.870$; \citealt{Marriage2011}). We generate mock observations using the observational setup introduced in Table~\ref{tab:ska_params}, but consider different SKA-MID configurations. The resulting comparison is shown in Fig.~\ref{fig:elgordo}. The comparison between the mock tSZ images generated using the nominal AA* and AA4 setups (i.e., only comprising the SKA subarray) clearly highlights the significant improvement that will be offered by the AA4 Band 5b integration. 

\section{Discussion}\label{sec:discussion}

\subsection{Comparison to AA*}

The mass-redshift thresholds explored in Section~\ref{sec:sim:massz} and displayed in Figure~\ref{fig:masszeta} clearly display the importance of the full SKA-MID configuration for tSZ studies with the SKA. A similar case can be observed in the comparison of Figure~\ref{fig:elgordo}, demonstrating the fundamental impact of the improved spatial coverage and sensitivity of the AA4 configuration over AA* for achieving high-fidelity SZ imaging. Although some limited SZ science may be done with AA*, unlocking the SKA's potential as an effective high-resolution tSZ instrument relies crucially on the increased density of short baselines offered by the full AA4 configuration. 

\subsection{Inclusion of MeerKAT antennas}\label{sec:discussion:meerkat}

\begin{figure}
    \centering
    \includegraphics[width=0.49\linewidth]{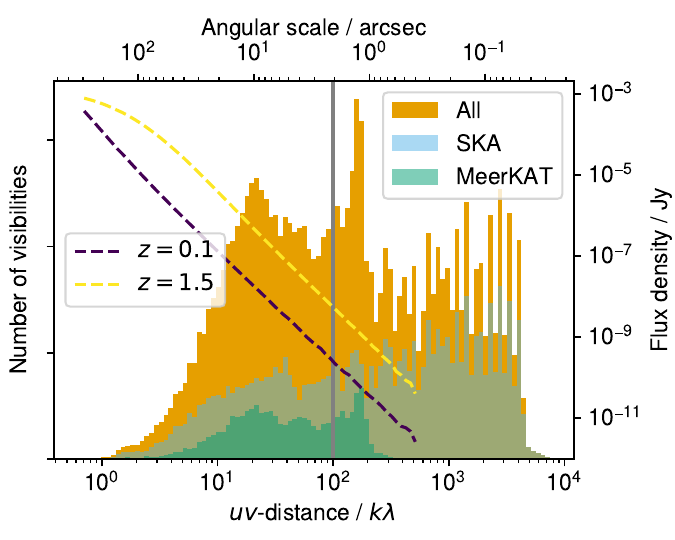}
    \includegraphics[width=0.49\linewidth]{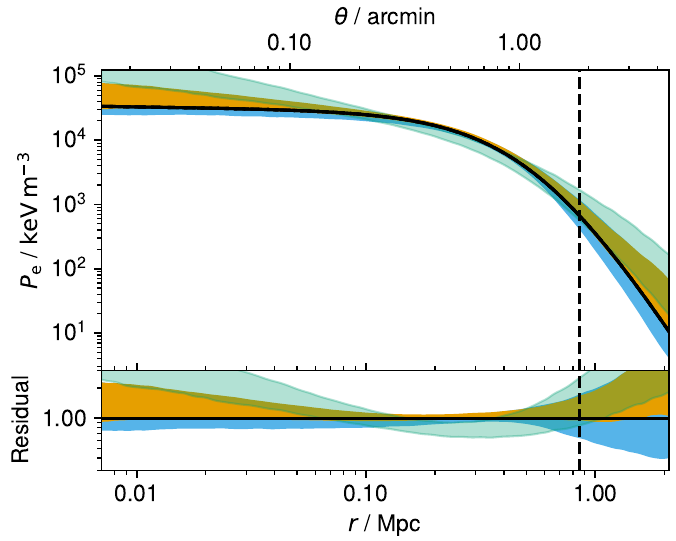}
    \caption{Left: baseline distributions for a 10 hour observation in AA4 configuration, including SKA-MID antennas only, MeerKAT antennas only, and all baselines (Band 5bii; histograms, left-hand axis).  We note that the `all baselines' observation includes SKA-only, MeerKAT-only and SKA $\times$ MeerKAT baselines and therefore contains more visibilities than the sum of the SKA-only and MeerKAT-only observations.  Dashed lines show example A10 cluster signal profiles in 5bii, with $M_{500} = 5 \times 10^{14} M_\odot$ and redshift as indicated in the legend.  The vertical line indicates the $uv$-range used for analyses. Right: comparison between pressure profile constraints on the \citet{2014ApJ...794...67M} non-cool-core average cluster, based on a 10-hour observation using the AA4 subarray configurations, using the same colour scheme as the left-hand plot.}
    \label{fig:uvamps_hybrid}
\end{figure}

Although not part of the main development roadmap for SKA-MID, funding dedicated to the Band 5b upgrade of MeerKAT has recently been secured by the Italian Institute for Astrophysics. The left-hand plot in Figure~\ref{fig:uvamps_hybrid} illustrates the SKA-MID AA4 baseline distribution with and without the MeerKAT antennas.  It is clear that the inclusion of MeerKAT would greatly increase the sampling of $uv$-distances relevant to cluster scales, increasing not only sensitivity but image fidelity.

As seen in Section~\ref{sec:sim:massz}, the enhanced large-scale sensitivity of MeerKAT compared to the SKA array will provide an immediate boost compared to SKA-MID AA*, bringing forward the capability to detect and characterize the tSZ signal from intermediate-mass systems prior to the full SKA-MID deployment. However, it is the full correlation of the two subarrays and the corresponding increase of short and intermediate baselines that will unlock the full capabilities of SKA-MID in advancing tSZ science, enabling high-fidelity imaging of the ICM even for galaxy clusters at $z=2$ with a mass as low as $M_{500}\simeq5\times10^{13}~\mathrm{M_{\odot}}$.

We further explore the potential for improvement with MeerKAT antennas included by repeating the pressure profile analysis from Section~\ref{sec:sim:profiles}.  The right-hand plot in Figure~\ref{fig:uvamps_hybrid} shows the improvement in cluster pressure profile constraints when the MeerKAT antennas are included.  The cluster is analysed under scenario two (uninformative priors); it can be seen that the pressure profile reconstruction is improved across the whole radial range when MeerKAT antennas are included, and particularly in the outer region where the extra large-scale information improves the reconstruction significantly. A qualitative comparison of the improvement that will be brought in the context of resolved tSZ imaging by the integration of the MeerKAT subarray can also be found in Figure~\ref{fig:elgordo}.

\subsection{Higher frequency bands}

There are two prospective extensions to the baseline SKA-MID frequency configuration.  These are Band 5$+$ (22.5 -- 25\,GHz) and Band 6 (36.25 –- 38.75\,GHz; \citealt{beyondb5}).  In terms of tSZ effect observations, there will be a trade-off between the greater amplitude of tSZ signals at these higher frequencies, versus the higher anticipated noise levels and the increase in baseline lengths as measured in $\lambda$, resolving out more of the large scale structure.  We therefore anticipate that these higher-frequency bands will be of most benefit for observing small-scale substructures.

We illustrate the potential advantages of the higher-frequency bands by considering SZ observations of ICM cavities: circular regions in the ICM where there is a lack of X-ray emission, typically aligned with large-scale jets from the central active galactic nucleus (AGN).  These cavities are generally considered to be an indicator of AGN feedback injecting energy into the ICM and disrupting cooling flow activity, however there is an open question regarding the contents of the cavities which SZ observations can aid in answering.  SKA observations of ICM cavities have already been explored in \citet{2025PASA...42...31G}; here we showcase one particular cavity configuration that shows the advantage of higher-frequency observations.

\begin{figure}
    \centering
    \includegraphics[width=\linewidth]{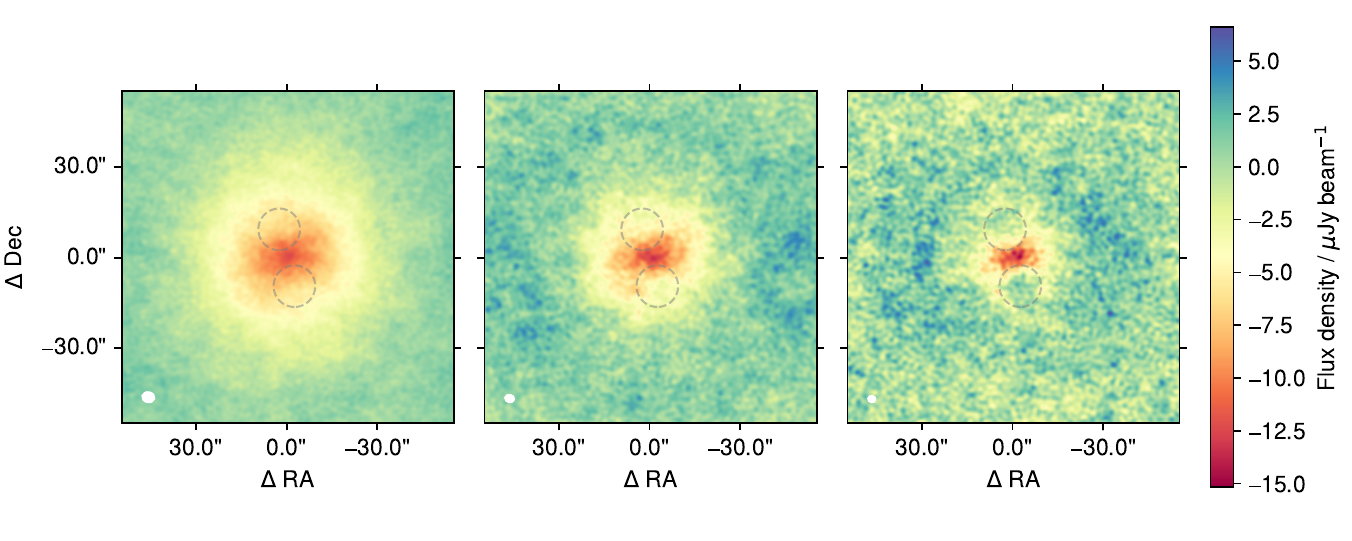}
    \caption{Example (dirty) images of ICM cavities in a $z=1.5$ cluster after two 10-hour observations, with Band 5bii (left), 5$+$ (centre) and 6 (right).  It can be seen that as the frequency increases, sensitivity to the global ICM decreases, but sensitivity to the smaller-scale cavities increases.  The model cavity positions are shown with dashed grey circles for comparison, and the synthesised beams are shown with white ellipses in the bottom left-hand corner of each plot.  These images have been made with visibilities with $\sqrt{u^2 + v^2} < 100$\,k$\lambda$.}
    \label{fig:cavities}
\end{figure}

We simulated a cluster of mass $M_{500} = 5 \times 10^{14} M_\odot$ and $z = 1.5$.  We then added a pair of cavities to the simulated $y$-map, with a radius of 60\,kpc (6.9\,arcsec) and a suppression factor of 0.99 compared to the global ICM signal \citep{2019ApJ...871..195A}.  We simulated observations at each of Bands 5bii, 5$+$ and 6, adding noise according to the \citet{braun2019anticipatedperformancesquarekilometre} noise model, rescaled to match the current sensitivity calculator at Band 5bii. Maps made from two simulated 10-hour observations are shown in Figure~\ref{fig:cavities} at each frequency band, showing the improved detection significance at higher frequency.  We also analysed the simulations at each frequency as in Section~\ref{sec:sim:profiles} both with and without including cavities in the fitted model, to obtain a detection significance estimate.  In Table~\ref{tab:cavities} we report the detection significances and constraints on the suppression factor, showing the utility of the higher-frequency bands for detecting these smaller-scale features.

\begin{table}
    \centering
    \begin{tabular}{lcc}
    \hline\hline
Frequency band & $\Delta\ln(\mathcal{Z})$ & $\sigma_f$ \\\hline
5bii & 23.2 (17.6 -- 35.0) & 0.34 (0.29 -- 0.45)\\
5$+$ & 27.2 (25.2 -- 38.5) & 0.29 (0.13 -- 0.35)\\
6 & 57.3 (25.8 -- 62.3) & 0.10 (0.015 -- 0.31)\\\hline
    \end{tabular}
    \caption{Results from the Bayesian $uv$-plane analysis of five realizations of two 10-hour simulated observations, at Bands 5bii, 5$+$ and 6.  The log-evidence difference $\Delta\ln(\mathcal{Z})$ is calculated for a cluster model with cavities with respect to a cluster model with no cavities; a value $>5$ generally indicates a strong detection.  $\sigma_f$ is the error on the suppression factor estimate, which indicates how useful the observation will be for constraining this factor, and hence for investigating the astrophysics of the cavity contents.  In each case we report the median from the 5 realizations, and the lower and upper bounds in brackets.}
    \label{tab:cavities}
\end{table}

\subsection{Synergies with other SKA frequency bands}\label{sec:discussion:synch}

As noted in our simulation sections, at Band 5b we expect there will be significant contamination to the tSZ signal from synchrotron radio signals, both compact (AGN, radio galaxies) and diffuse (cluster halos, relics, etc).  We expect that the longer baselines of the Band 5b observations will be very effective at removing the compact sources (see Fig.~\ref{fig:uvamps_A10}), however the diffuse sources, covering similar angular scales to the tSZ signal, will require a different treatment.

Since the spectral indices of these sources are steep (typically $\alpha < -1$, for $S \propto \nu^\alpha$), lower-frequency SKA observations will be much more sensitive to these synchrotron sources (and much less sensitive to the tSZ signal, which gets fainter at lower frequency with $\alpha \approx +2$, in the Rayleigh-Jeans regime).  In addition, lower-frequency observations with the SKA will have better sensitivity to large angular scales than Band 5b since the baselines (as measured in $\lambda$) will be shorter.  The combination of these two factors means that a short observation in Band 1 or 2, or even with SKA-LOW, will be sufficient to characterize the diffuse sources (see also \citealt{Cassano01.2026.SKA}, \citealt{deGasperin01.2026.SKA}, \citealt{Gitti01.2026.SKA}, \citealt{ArpanPal01.2026.SKA}, \citealt{Santra01.2026.SKA}, \citealt{Vacca01.2026.SKA}).  The diffuse sources can then be removed or jointly modelled with the SZ signal, making use also of the in-band spectral variation of the different types of sources within Band 5b.

As well as a challenge, this presents an opportunity: with the combination of frequency bands, the SKA will simultaneously probe the bulk thermal component and its substructures, tSZ fluctuations as a diagnostic of ICM turbulence, and the diffuse radio synchrotron emission associated with relativistic electrons and magnetic fields (see also \citealt{Akahori01.2026.SKA}, \citealt{Cuciti01.2026.SKA}, \citealt{Ignesti01.2026.SKA}, \citealt{Loi01.2026.SKA}, \citealt{OSullivan01.2026.SKA}). Since these components are intricately connected, SKA will offer a unique opportunity to study them together with a single instrument, down to low mass and up to high redshifts.

\subsection{Synergies with other instruments}
The capability of SKA-MID Band 5b to perform wide-field SZ imaging at high angular resolution at centimetre wavelengths will be uniquely complementary to observations from current and upcoming (sub)millimetre-wave facilities. The most straightforward aspect will be the extension of the spectral coverage over a frequency regime that is minimally biased by contributions from relativistic and kinematic SZ contributions, enhancing the spectral modelling of the thermal SZ signal in combination with information from large-aperture single-dish instruments such as MUSTANG2 \citep{2014JLTP..176..808D}, NIKA2 \citep{2018A&A...609A.115A} (and their prospective upgrades\footnote{\url{https://greenbankobservatory.org/portal/gbt/instruments/mustang-3/}}), TolTEC \citep{2020SPIE11453E..02W} and MUSCAT \citep{2024JATIS..10d5003T}, or interferometric instruments such as NOEMA \citep{2023pcsf.conf..308N}, ALMA \citep{2009IEEEP..97.1463W}, and its forthcoming WSU \citep{2023pcsf.conf..304C}. 

The wide extent of the field of view compared to the aforementioned facilities and the resulting mapping speed will also allow SKA-MID to serve as an efficient high-resolution counterpart to low-resolution wide-field SZ measurements available from survey experiments like the ongoing SPT-3G \citep{2014SPIE.9153E..1PB}, Simons Observatory \citep{SO2019,SOLAT2025}, and CCAT \citep{2023ApJS..264....7C}. These will be essential to integrate large-scale information and enable a high-dynamic range imaging and characterization of the SZ structure of galaxy clusters. On a longer timescale, the Atacama Large Aperture Submillimeter Survey (AtLAST; \citealt{2025A&A...694A.142M}) and CMB-HD \citep{2019BAAS...51g...6S} will deepen such a synergy by bridging the gap between high-resolution, small-field tSZ imaging and large-scale, low-resolution surveys, opening in combination with SKA-MID to a high-sensitivity, multi-frequency coverage of the radio and SZ signature of galaxy clusters.

On the X-ray side, the combination of SKA-MID tSZ measurements with data from existing telescope as e.g. \textit{Chandra} \citep{2002PASP..114....1W}, XMM–Newton \citep{2001A&A...365L...1J}, eROSITA \citep{2021A&A...647A...1P}, or XRISM \citep{2024PASJ...76.1186X} will enable a powerful joint inference of the thermodynamic and kinematic properties of the ICM up to $z\sim2$; see also \citet{Kurahara01.2026.SKA} for a detailed exploration of SKA-XRISM synergies. Future missions such as NewAthena \citep{newathena25}, AXIS \citep{AXIS2025}, LEM \citep{2023JATIS...9d1008P}, and HUBS \citep{2020JLTP..199..502C} will introduce significant improvements in terms of effective area, field of view, energy and/or spatial resolution, further enhancing the possibility of achieving a thorough characterization of the ICM physical state.

\section{Conclusion}

In this Chapter, we have demonstrated that the SKA will be a powerful tool for tSZ observations.  Observing in Band 5b, the compact core of antennas provides sensitivity to the large angular scales necessary for observations of the global intracluster medium, while the longer baselines provide a high-resolution tSZ view of cluster substructures.  These characteristics will enable the SKA to detect clusters at masses and redshifts competitive with other current and near-future SZ instruments; constrain cluster profiles with a precision equivalent to or better than X-ray observations; detect and constrain pressure fluctuations as a tracer of turbulence; and detect intracluster medium shocks.

We have also demonstrated that the SKA would be even more powerful if the MeerKAT antennas were equipped with Band 5 receivers.  This would boost the number of short baselines, dramatically increasing the SKA's sensitivity to cluster-scale emission.  Another improvement could be made by observing at the proposed higher frequency bands, 5$+$ and Band 6, where the higher SZ signal amplitude provides better sensitivity to smaller-scale cluster features such as cavities and shocks.

The SKA's tSZ observations will be highly complementary to observations at other frequencies with the SKA, as well as observations in other wavebands with current and future instruments.  Combining high-resolution, high-sensitivity observations of clusters at wavelengths from radio through to X-ray will open up new horizons for thoroughly understanding the astrophysics of clusters and the thermodynamics of the intracluster medium.

\section*{Acknowledgements}

We thank the anonymous referee for helpful suggestions which improved the presentation of this Chapter.  

We acknowledge the use of the Rāpoi high-performance computing facility of Te Herenga Waka--Victoria University of Wellington.

This work was supported by the French government through the France 2030 investment plan managed by the National Research Agency (ANR), as part of the Initiative of Excellence of Universit\'e C\^ote d'Azur under reference number ANR-15-IDEX-01. 

This work was supported by the French government, through the UCA$^{\rm J.E.D.I.}$ Investments in the Future project managed by the National Research Agency (ANR) with the reference number ANR-15-IDEX-01.

\bibliographystyle{abbrvnat-maxbibnames4}
\bibliography{chapter}

@misc{braun2019anticipatedperformancesquarekilometre,
      title={Anticipated Performance of the {Square Kilometre Array} -- Phase 1 {(SKA1)}}, 
      author={Robert Braun and Anna Bonaldi and Tyler Bourke and Evan Keane and Jeff Wagg},
      year={2019},
      eprint={1912.12699},
      archivePrefix={arXiv},
      primaryClass={astro-ph.IM},
      url={https://arxiv.org/abs/1912.12699}, 
}

@TECHREPORT{beyondb5,
  AUTHOR =        {{Conway}, J. and {Beswick}, R. and {Bourke}, T. and {Coriat}, M. and {Ferrari}, C. and {Jimenez-Serra}, I. and {Muller}, S. and {Sargent}, M.},
  TITLE =         {{SKA1} Beyond {15GHz}:
The Science case for Band 6},
  YEAR  =         {2020},
  URL   =         {https://www.skao.int/sites/default/files/documents/d38-ScienceCase_band6_Feb2020.pdf}
}

@ARTICLE{ACT2025,
	journal={The Open Journal of Astrophysics},
	doi={10.33232/001c.155863},
	publisher={Maynooth Academic Publishing},
	title={The Atacama Cosmology Telescope: DR6 Sunyaev-Zel'dovich Selected Galaxy Clusters Catalog},
	volume=9,
	author={{ACT/DES/HSC Collaboration}},
	date={2026-01-27},
	year=2026,
	month=Jan,
	day=27,
          eid = {arXiv:2507.21459},
        pages = {arXiv:2507.21459},
archivePrefix = {arXiv},
       eprint = {2507.21459},
 primaryClass = {astro-ph.CO},
       adsurl = {https://doi.org/10.33232/001c.155863}
}

@ARTICLE{Bleem2020,
       author = {{Bleem}, L.~E. and {Bocquet}, S. and {Stalder}, B. and {Gladders}, M.~D. and {Ade}, P.~A.~R. and {Allen}, S.~W. and {Anderson}, A.~J. and {Annis}, J. and {Ashby}, M.~L.~N. and {Austermann}, J.~E. and {Avila}, S. and {Avva}, J.~S. and {Bayliss}, M. and {Beall}, J.~A. and {Bechtol}, K. and {Bender}, A.~N. and {Benson}, B.~A. and {Bertin}, E. and {Bianchini}, F. and {Blake}, C. and {Brodwin}, M. and {Brooks}, D. and {Buckley-Geer}, E. and {Burke}, D.~L. and {Carlstrom}, J.~E. and {Rosell}, A. Carnero and {Carrasco Kind}, M. and {Carretero}, J. and {Chang}, C.~L. and {Chiang}, H.~C. and {Citron}, R. and {Moran}, C. Corbett and {Costanzi}, M. and {Crawford}, T.~M. and {Crites}, A.~T. and {da Costa}, L.~N. and {de Haan}, T. and {De Vicente}, J. and {Desai}, S. and {Diehl}, H.~T. and {Dietrich}, J.~P. and {Dobbs}, M.~A. and {Eifler}, T.~F. and {Everett}, W. and {Flaugher}, B. and {Floyd}, B. and {Frieman}, J. and {Gallicchio}, J. and {Garc{\'\i}a-Bellido}, J. and {George}, E.~M. and {Gerdes}, D.~W. and {Gilbert}, A. and {Gruen}, D. and {Gruendl}, R.~A. and {Gschwend}, J. and {Gupta}, N. and {Gutierrez}, G. and {Halverson}, N.~W. and {Harrington}, N. and {Henning}, J.~W. and {Heymans}, C. and {Holder}, G.~P. and {Hollowood}, D.~L. and {Holzapfel}, W.~L. and {Honscheid}, K. and {Hrubes}, J.~D. and {Huang}, N. and {Hubmayr}, J. and {Irwin}, K.~D. and {James}, D.~J. and {Jeltema}, T. and {Joudaki}, S. and {Khullar}, G. and {Klein}, M. and {Knox}, L. and {Kuropatkin}, N. and {Lee}, A.~T. and {Li}, D. and {Lidman}, C. and {Lowitz}, A. and {MacCrann}, N. and {Mahler}, G. and {Maia}, M.~A.~G. and {Marshall}, J.~L. and {McDonald}, M. and {McMahon}, J.~J. and {Melchior}, P. and {Menanteau}, F. and {Meyer}, S.~S. and {Miquel}, R. and {Mocanu}, L.~M. and {Mohr}, J.~J. and {Montgomery}, J. and {Nadolski}, A. and {Natoli}, T. and {Nibarger}, J.~P. and {Noble}, G. and {Novosad}, V. and {Padin}, S. and {Palmese}, A. and {Parkinson}, D. and {Patil}, S. and {Paz-Chinch{\'o}n}, F. and {Plazas}, A.~A. and {Pryke}, C. and {Ramachandra}, N.~S. and {Reichardt}, C.~L. and {Remolina Gonz{\'a}lez}, J.~D. and {Romer}, A.~K. and {Roodman}, A. and {Ruhl}, J.~E. and {Rykoff}, E.~S. and {Saliwanchik}, B.~R. and {Sanchez}, E. and {Saro}, A. and {Sayre}, J.~T. and {Schaffer}, K.~K. and {Schrabback}, T. and {Serrano}, S. and {Sharon}, K. and {Sievers}, C. and {Smecher}, G. and {Smith}, M. and {Soares-Santos}, M. and {Stark}, A.~A. and {Story}, K.~T. and {Suchyta}, E. and {Tarle}, G. and {Tucker}, C. and {Vanderlinde}, K. and {Veach}, T. and {Vieira}, J.~D. and {Wang}, G. and {Weller}, J. and {Whitehorn}, N. and {Wu}, W.~L.~K. and {Yefremenko}, V. and {Zhang}, Y.},
        title = "{The SPTpol Extended Cluster Survey}",
      journal = {\apjs},
         year = 2020,
        month = mar,
       volume = {247},
       number = {1},
          eid = {25},
        pages = {25},
          doi = {10.3847/1538-4365/ab6993},
archivePrefix = {arXiv},
       eprint = {1910.04121},
 primaryClass = {astro-ph.CO},
       adsurl = {https://ui.adsabs.harvard.edu/abs/2020ApJS..247...25B}
}

@ARTICLE{Bleem2023,
	journal={The Open Journal of Astrophysics},
	doi={10.21105/astro.2311.07512},
	publisher={Maynooth Academic Publishing},
	title={Galaxy Clusters Discovered via the Thermal Sunyaev-Zel’dovich Effect in the 500-square-degree SPTpol Survey},
	volume=7,
	author={Bleem, L.E. and Klein, M. and Abbott, T. M. C. and Ade, P. A. R. and Aguena, M. and Alves, O. and Anderson, A. J. and Andrade-Oliveira, F. and Ansarinejad, B. and Archipley, M. and Ashby, M. L. N. and Austermann, J. E. and Bacon, D. and Beall, J. A. and Bender, A. N. and Benson, B. A. and Bianchini, F. and Bocquet, S. and Brooks, D. and Burke, D. L. and Calzadilla, M. and Carlstrom, J. E. and Rosell, A. Carnero and Carretero, J. and Chang, C. L. and Chaubal, P. and Chiang, H. C. and Chou, T-L. and Citron, R. and Moran, C. Corbett and Costanzi, M. and Crawford, T. M. and Crites, A. T. and da Costa, L. N. and de Haan, T. and De Vicente, J. and Desai, S. and Dobbs, M. A. and Doel, P. and Everett, W. and Ferrero, I. and Flaugher, B. and Floyd, B. and Friedel, D. and Frieman, J. and Gallicchio, J. and Garc'ia-Bellido, J. and Gatti, M. and George, E. M. and Giannini, G. and Grandis, S. and Gruen, D. and Gruendl, R. A. and Gupta, N. and Gutierrez, G. and Halverson, N. W. and Hinton, S. R. and Holder, G. P. and Hollowood, D. L. and Holzapfel, W. L. and Honscheid, K. and Hrubes, J. D. and Huang, N. and Hubmayr, J. and Irwin, K. D. and Mena-Fernández, J. and James, D. J. and Kéruzoré, F. and Knox, L. and Kuehn, K. and Lahav, O. and Lee, A. T. and Lee, S. and Li, D. and Lowitz, A. and Marshal, J. L. and McDonald, M. and McMahon, J. J. and Menanteau, F. and Meyer, S. S. and Miquel, R. and Mohr, J. J. and Montgomery, J. and Myles, J. and Natoli, T. and Nibarger, J. P. and Noble, G. I. and Novosad, V. and Ogando, R. L. C. and Padin, S. and Patil, S. and Pereira, M. E. S. and Pieres, A. and Malag'on, A. A. Plazas and Pryke, C. and Reichardt, C. L. and Rodr'iguez-Monroy, M. and Romer, A. K. and Ruhl, J. E. and Saliwanchik, B. R. and Salvati, L. and Sanchez, E. and Saro, A. and Schaffer, K. K. and Schrabback, T. and Sevilla-Noarbe, I. and Sievers, C. and Smecher, G. and Smith, M. and Somboonpanyakul, T. and Stalder, B. and Stark, A. A. and Suchyta, E. and Swanson, M. E. C. and Tarle, G. and To, C. and Tucker, C. and Veach, T. and Vieira, J. D. and Vincenzi, M. and Wang, G. and Weller, J. and Whitehorn, N. and Wiseman, P. and Wu, W. L. K. and Yefremenko, V. and Zebrowski, J. A. and Zhang, Y.},
	date={2024-02-09},
	year=2024,
	month=feb,
	day=9,
          eid = {arXiv:2311.07512},
        pages = {arXiv:2311.07512},
archivePrefix = {arXiv},
       eprint = {2311.07512},
 primaryClass = {astro-ph.CO},
       adsurl = {https://doi.org/10.21105/astro.2311.07512}
}

@ARTICLE{Broekema2015,
       author = {{Broekema}, P.~C. and {van Nieuwpoort}, R.~V. and {Bal}, H.~E.},
        title = "{The Square Kilometre Array Science Data Processor. Preliminary compute platform design}",
      journal = {Journal of Instrumentation},
         year = 2015,
        month = jul,
       volume = {10},
       number = {7},
          eid = {C07004},
        pages = {C07004},
          doi = {10.1088/1748-0221/10/07/C07004},
       adsurl = {https://ui.adsabs.harvard.edu/abs/2015JInst..10C7004B}
}

@article{Brunetti2014,
  author    = {Brunetti, G. and Jones, T.},
  title     = {Cosmic Rays in Galaxy Clusters and Their Non-Thermal Emission},
  journal   = {International Journal of Modern Physics D},
  year      = {2014},
  volume    = {23},
  number    = {4},
  pages     = {1430007},
  doi       = {10.1142/S0218271814300079},
archivePrefix = {arXiv},
       eprint = {1401.7519},
 primaryClass = {astro-ph.CO},
       adsurl = {https://arxiv.org/abs/1401.7519}
}

@article{Carlstrom2002,
  author    = {Carlstrom, J. and Holder, G. and Reese, E.},
  title     = {Cosmology with the Sunyaev-Zel’dovich Effect},
  journal   = {Annual Review of Astronomy and Astrophysics},
  year      = {2002},
  volume    = {40},
  pages     = {643--680},
  doi       = {10.1146/annurev.astro.40.060401.093803},
archivePrefix = {arXiv},
 primaryClass = {astro-ph.CO},
       adsurl = {https://arxiv.org/abs/astro-ph/0208192}
}

@INPROCEEDINGS{Farnes2018,
  author={Farnes, Jamie S. and Mort, Ben and Dulwich, Fred and Adámek, Karel and Brown, Anna and Novotny, Jan and Salvini, Stef and Armour, Wes},
  booktitle={2018 IEEE 14th International Conference on e-Science (e-Science)}, 
  title={Building the World's Largest Radio Telescope: The Square Kilometre Array Science Data Processor}, 
  year={2018},
  volume={},
  number={},
  pages={366-367},
  doi={10.1109/eScience.2018.00101}
}

@article{Kitayama2016,
  author = {{Kitayama}, Tetsu and {Ueda}, Shutaro and {Takakuwa}, Shigehisa and {Tsutsumi}, Takahiro and {Komatsu}, Eiichiro and {Akahori}, Takuya and {Iono}, Daisuke and {Izumi}, Takuma and {Kawabe}, Ryohei and {Kohno}, Kotaro and {Matsuo}, Hiroshi and {Ota}, Naomi and {Suto}, Yasushi and {Takizawa}, Motozaku and {Yoshikawa}, Kohji},
        title = "{The Sunyaev-Zel'dovich effect at 5″: RX J1347.5-1145 imaged by ALMA}",
      journal = {\pasj},
         year = 2016,
        month = oct,
       volume = {68},
       number = {5},
          eid = {88},
        pages = {88},
          doi = {10.1093/pasj/psw082},
archivePrefix = {arXiv},
       eprint = {1607.08833},
 primaryClass = {astro-ph.CO},
       adsurl = {https://ui.adsabs.harvard.edu/abs/2016PASJ...68...88K}
}

@ARTICLE{Kornoelje2025,
       author = {{Kornoelje}, K. and {Bleem}, L.~E. and {Rykoff}, E.~S. and {Abbott}, T.~M.~C. and {Ade}, P.~A.~R. and {Aguena}, M. and {Alves}, O. and {Anderson}, A.~J. and {Andrade-Oliveira}, F. and {Ansarinejad}, B. and {Archipley}, M. and {Ashby}, M.~L.~N. and {Austermann}, J.~E. and {Bacon}, D. and {Balkenhol}, L. and {Beall}, J.~A. and {Benabed}, K. and {Bender}, A.~N. and {Benson}, B.~A. and {Bianchini}, F. and {Bocquet}, S. and {Bouchet}, F.~R. and {Brooks}, D. and {Burke}, D.~L. and {Calzadilla}, M. and {Camphuis}, E. and {Carlstrom}, J.~E. and {Carnero Rosell}, A. and {Carretero}, J. and {Chang}, C.~L. and {Chaubal}, P. and {Chiang}, H.~C. and {Chichura}, P.~M. and {Chokshi}, A. and {Chou}, T. -L. and {Citron}, R. and {Coerver}, A. and {Corbett Moran}, C. and {Costanzi}, M. and {Crawford}, T.~M. and {Crites}, A.~T. and {da Costa}, L.~N. and {Daley}, C. and {de Haan}, T. and {De Vicente}, J. and {Desai}, S. and {Dibert}, K.~R. and {Dobbs}, M.~A. and {Doel}, P. and {Doohan}, M. and {Doussot}, A. and {Dutcher}, D. and {Everett}, W. and {Everett}, S. and {Feng}, C. and {Ferguson}, K.~R. and {Ferrero}, I. and {Fichman}, K. and {Flaugher}, B. and {Floyd}, B. and {Foster}, A. and {Friedel}, D. and {Frieman}, J. and {Galli}, S. and {Gallicchio}, J. and {Gambrel}, A.~E. and {Garc'ia-Bellido}, J. and {Gardner}, R.~W. and {Gassis}, R. and {Gatti}, M. and {Ge}, F. and {George}, E.~M. and {Giannini}, G. and {Goeckner-Wald}, N. and {Grandis}, S. and {Gruen}, D. and {Gruendl}, R.~A. and {Gualtieri}, R. and {Guidi}, F. and {Mahler}, Guillaume and {Guns}, S. and {Gupta}, N. and {Gutierrez}, G. and {Halverson}, N.~W. and {Hinton}, S.~R. and {Hivon}, E. and {Holder}, G.~P. and {Hollowood}, D.~L. and {Holzapfel}, W.~L. and {Honscheid}, K. and {Hood}, J.~C. and {Hrubes}, J.~D. and {Hryciuk}, A. and {Huang}, N. and {Hubmayr}, J. and {Irwin}, K.~D. and {Mena-Fern\textbackslash'andez}, J. and {James}, D.~J. and {K\textbackslash'eruzor\textbackslash'e}, F. and {Khalife}, A.~R. and {Klein}, M. and {Knox}, L. and {Korman}, M. and {Kuehn}, K. and {Kuo}, C. -L. and {Lahav}, O. and {Lee}, A.~T. and {Lee}, S. and {Levy}, K. and {Li}, D. and {Lima}, M. and {Lowitz}, A.~E. and {Lowitz}, A. and {Lu}, C. and {Maniyar}, A. and {Marshal}, J.~L. and {Marshall}, J.~L. and {Martsen}, E.~S. and {Bayliss}, Matthew B. and {McDonald}, M. and {McMahon}, J.~J. and {Menanteau}, F. and {Millea}, M. and {Miquel}, R. and {Mohr}, J.~J. and {Montgomery}, J. and {Myles}, J. and {Nakato}, Y. and {Natoli}, T. and {Nibarger}, J.~P. and {Noble}, G.~I. and {Novosad}, V. and {Ogando}, R.~L.~C. and {Omori}, Y. and {Ouellette}, A. and {Padin}, S. and {Pan}, Z. and {Patil}, S. and {Pereira}, M.~E.~S. and {Phadke}, K.~A. and {Pieres}, A. and {Plazas Malag'on}, A.~A. and {Pollak}, A.~W. and {Prabhu}, K. and {Pryke}, C. and {Quan}, W. and {Raghunathan}, S. and {Rahimi}, M. and {Rahlin}, A. and {Reichardt}, C.~L. and {Rodr'iguez-Monroy}, M. and {Romer}, A.~K. and {Rouble}, M. and {Ruhl}, J.~E. and {Saliwanchik}, B.~R. and {Salvati}, L. and {Samuroff}, S. and {Sanchez}, E. and {Saro}, A. and {Schaffer}, K.~K. and {Schiappucci}, E. and {Schrabback}, T. and {Sevilla-Noarbe}, I. and {Sievers}, C. and {Smecher}, G. and {Smith}, M. and {Sobrin}, J.~A. and {Somboonpanyakul}, T. and {Stalder}, B. and {Stark}, A.~A. and {Suchyta}, E. and {Swanson}, M.~E.~C. and {Tandoi}, C. and {Tarle}, G. and {Thorne}, B. and {To}, C. and {Trendafilova}, C. and {Tucker}, C. and {Umilta}, C. and {Veach}, T. and {Vieira}, J.~D. and {Vincenzi}, M. and {Vitrier}, A. and {Wan}, Y. and {Wang}, G. and {Weaverdyck}, N. and {Weller}, J. and {Whitehorn}, N. and {Wiseman}, P. and {Wu}, W.~L.~K. and {Yefremenko}, V. and {Young}, M.~R. and {Zebrowski}, J.~A. and {Zhang}, Y.},
        title = "{The SPT-Deep Cluster Catalog: Sunyaev-Zel'dovich Selected Clusters from Combined SPT-3G and SPTpol Measurements over 100 Square Degrees}",
      journal = {arXiv e-prints},
         year = 2025,
        month = mar,
          eid = {arXiv:2503.17271},
        pages = {arXiv:2503.17271},
          doi = {10.48550/arXiv.2503.17271},
archivePrefix = {arXiv},
       eprint = {2503.17271},
 primaryClass = {astro-ph.CO},
       adsurl = {https://ui.adsabs.harvard.edu/abs/2025arXiv250317271K}
}

@ARTICLE{Planck2016,
       author = {{Planck Collaboration}},
        title = "{Planck 2015 results. XXVII. The second Planck catalogue of Sunyaev-Zeldovich sources}",
      journal = {\aap},
         year = 2016,
        month = sep,
       volume = {594},
          eid = {A27},
        pages = {A27},
          doi = {10.1051/0004-6361/201525823},
archivePrefix = {arXiv},
       eprint = {1502.01598},
 primaryClass = {astro-ph.CO},
       adsurl = {https://ui.adsabs.harvard.edu/abs/2016A&A...594A..27P}
}

@ARTICLE{Planck2016b,
       author = {{Planck Collaboration}},
        title = "{Planck 2015 results. XXIV. Cosmology from Sunyaev-Zeldovich cluster counts}",
      journal = {\aap},
         year = 2016,
        month = sep,
       volume = {594},
          eid = {A24},
        pages = {A24},
          doi = {10.1051/0004-6361/201525833},
archivePrefix = {arXiv},
       eprint = {1502.01597},
 primaryClass = {astro-ph.CO},
       adsurl = {https://ui.adsabs.harvard.edu/abs/2016A&A...594A..24P}
}

@ARTICLE{Arnaud2010,
       author = {{Arnaud}, M. and {Pratt}, G.~W. and {Piffaretti}, R. and {B{\"o}hringer}, H. and {Croston}, J.~H. and {Pointecouteau}, E.},
        title = "{The universal galaxy cluster pressure profile from a representative sample of nearby systems (REXCESS) and the Y$_{SZ}$ - M$_{500}$ relation}",
      journal = {\aap},
         year = 2010,
        month = jul,
       volume = {517},
          eid = {A92},
        pages = {A92},
          doi = {10.1051/0004-6361/200913416},
archivePrefix = {arXiv},
       eprint = {0910.1234},
 primaryClass = {astro-ph.CO},
       adsurl = {https://ui.adsabs.harvard.edu/abs/2010A&A...517A..92A}
}

@ARTICLE{2007ApJ...668....1N,
       author = {{Nagai}, Daisuke and {Kravtsov}, Andrey V. and {Vikhlinin}, Alexey},
        title = "{Effects of Galaxy Formation on Thermodynamics of the Intracluster Medium}",
      journal = {\apj},
         year = 2007,
        month = oct,
       volume = {668},
       number = {1},
        pages = {1-14},
          doi = {10.1086/521328},
archivePrefix = {arXiv},
       eprint = {astro-ph/0703661},
 primaryClass = {astro-ph},
       adsurl = {https://ui.adsabs.harvard.edu/abs/2007ApJ...668....1N}
}

@ARTICLE{2014ApJ...794...67M,
       author = {{McDonald}, M. and {Benson}, B.~A. and {Vikhlinin}, A. and {Aird}, K.~A. and {Allen}, S.~W. and {Bautz}, M. and {Bayliss}, M. and {Bleem}, L.~E. and {Bocquet}, S. and {Brodwin}, M. and {Carlstrom}, J.~E. and {Chang}, C.~L. and {Cho}, H.~M. and {Clocchiatti}, A. and {Crawford}, T.~M. and {Crites}, A.~T. and {de Haan}, T. and {Dobbs}, M.~A. and {Foley}, R.~J. and {Forman}, W.~R. and {George}, E.~M. and {Gladders}, M.~D. and {Gonzalez}, A.~H. and {Halverson}, N.~W. and {Hlavacek-Larrondo}, J. and {Holder}, G.~P. and {Holzapfel}, W.~L. and {Hrubes}, J.~D. and {Jones}, C. and {Keisler}, R. and {Knox}, L. and {Lee}, A.~T. and {Leitch}, E.~M. and {Liu}, J. and {Lueker}, M. and {Luong-Van}, D. and {Mantz}, A. and {Marrone}, D.~P. and {McMahon}, J.~J. and {Meyer}, S.~S. and {Miller}, E.~D. and {Mocanu}, L. and {Mohr}, J.~J. and {Murray}, S.~S. and {Padin}, S. and {Pryke}, C. and {Reichardt}, C.~L. and {Rest}, A. and {Ruhl}, J.~E. and {Saliwanchik}, B.~R. and {Saro}, A. and {Sayre}, J.~T. and {Schaffer}, K.~K. and {Shirokoff}, E. and {Spieler}, H.~G. and {Stalder}, B. and {Stanford}, S.~A. and {Staniszewski}, Z. and {Stark}, A.~A. and {Story}, K.~T. and {Stubbs}, C.~W. and {Vanderlinde}, K. and {Vieira}, J.~D. and {Williamson}, R. and {Zahn}, O. and {Zenteno}, A.},
        title = "{The Redshift Evolution of the Mean Temperature, Pressure, and Entropy Profiles in 80 SPT-Selected Galaxy Clusters}",
      journal = {\apj},
         year = 2014,
        month = oct,
       volume = {794},
       number = {1},
          eid = {67},
        pages = {67},
          doi = {10.1088/0004-637X/794/1/67},
archivePrefix = {arXiv},
       eprint = {1404.6250},
 primaryClass = {astro-ph.HE},
       adsurl = {https://ui.adsabs.harvard.edu/abs/2014ApJ...794...67M}
}

@article{Sarazin1985,
  author    = {C. L. Sarazin},
  title     = {X-ray Emission from Clusters of Galaxies},
  journal   = {Reviews of Modern Physics},
  year      = {1985},
  volume    = {58},
  pages     = {1-115},
  doi       = {10.1103/RevModPhys.58.1},
archivePrefix = {arXiv},
 primaryClass = {astro-ph.HE},
       adsurl = {https://ui.adsabs.harvard.edu/abs/1986RvMP...58....1S/abstract}
}

@ARTICLE{Vazza2016,
       author = {{Vazza}, Franco and {Wittor}, Denis and {Br{\"u}ggen}, Marcus and {Gheller}, Claudio},
        title = "{On the Non-Thermal Energy Content of Cosmic Structures}",
      journal = {Galaxies},
         year = 2016,
        month = nov,
       volume = {4},
       number = {4},
          eid = {60},
        pages = {60},
          doi = {10.3390/galaxies4040060},
archivePrefix = {arXiv},
       eprint = {1610.00129},
 primaryClass = {astro-ph.CO},
       adsurl = {https://ui.adsabs.harvard.edu/abs/2016Galax...4...60V}
}

@ARTICLE{Angelinelli2020,
       author = {{Angelinelli}, M. and {Vazza}, F. and {Giocoli}, C. and {Ettori}, S. and {Jones}, T.~W. and {Brunetti}, G. and {Br{\"u}ggen}, M. and {Eckert}, D.},
        title = "{Turbulent pressure support and hydrostatic mass bias in the intracluster medium}",
      journal = {\mnras},
         year = 2020,
        month = jun,
       volume = {495},
       number = {1},
        pages = {864-885},
          doi = {10.1093/mnras/staa975},
archivePrefix = {arXiv},
       eprint = {1905.04896},
 primaryClass = {astro-ph.CO},
       adsurl = {https://ui.adsabs.harvard.edu/abs/2020MNRAS.495..864A}
}

@ARTICLE{vanWeeren2019,
       author = {{van Weeren}, R.~J. and {de Gasperin}, F. and {Akamatsu}, H. and {Br{\"u}ggen}, M. and {Feretti}, L. and {Kang}, H. and {Stroe}, A. and {Zandanel}, F.},
        title = "{Diffuse Radio Emission from Galaxy Clusters}",
      journal = {\ssr},
         year = 2019,
        month = feb,
       volume = {215},
       number = {1},
          eid = {16},
        pages = {16},
          doi = {10.1007/s11214-019-0584-z},
archivePrefix = {arXiv},
       eprint = {1901.04496},
 primaryClass = {astro-ph.HE},
       adsurl = {https://ui.adsabs.harvard.edu/abs/2019SSRv..215...16V}
}

@ARTICLE{Dupourque2024,
       author = {{Dupourqu{\'e}}, S. and {Clerc}, N. and {Pointecouteau}, E. and {Eckert}, D. and {Gaspari}, M. and {Lovisari}, L. and {Pratt}, G.~W. and {Rasia}, E. and {Rossetti}, M. and {Vazza}, F. and {Balboni}, M. and {Bartalucci}, I. and {Bourdin}, H. and {De Luca}, F. and {De Petris}, M. and {Ettori}, S. and {Ghizzardi}, S. and {Mazzotta}, P.},
        title = "{CHEX-MATE: Turbulence in the intra-cluster medium from X-ray surface brightness fluctuations}",
      journal = {\aap},
         year = 2024,
        month = jul,
       volume = {687},
          eid = {A58},
        pages = {A58},
          doi = {10.1051/0004-6361/202348701},
archivePrefix = {arXiv},
       eprint = {2403.03064},
 primaryClass = {astro-ph.CO},
       adsurl = {https://ui.adsabs.harvard.edu/abs/2024A&A...687A..58D}
}

@ARTICLE{Heinrich2024,
       author = {{Heinrich}, Annie and {Zhuravleva}, Irina and {Zhang}, Congyao and {Churazov}, Eugene and {Forman}, William and {van Weeren}, Reinout J.},
        title = "{Merger-driven multiscale ICM density perturbations: testing cosmological simulations and constraining plasma physics}",
      journal = {\mnras},
         year = 2024,
        month = mar,
       volume = {528},
       number = {4},
        pages = {7274-7299},
          doi = {10.1093/mnras/stae208},
archivePrefix = {arXiv},
       eprint = {2401.15179},
 primaryClass = {astro-ph.HE},
       adsurl = {https://ui.adsabs.harvard.edu/abs/2024MNRAS.528.7274H}
}

@ARTICLE{Khatri2016,
       author = {{Khatri}, Rishi and {Gaspari}, Massimo},
        title = "{Thermal SZ fluctuations in the ICM: probing turbulence and thermodynamics in Coma cluster with Planck}",
      journal = {\mnras},
         year = 2016,
        month = nov,
       volume = {463},
       number = {1},
        pages = {655-669},
          doi = {10.1093/mnras/stw2027},
archivePrefix = {arXiv},
       eprint = {1604.03106},
 primaryClass = {astro-ph.CO},
       adsurl = {https://ui.adsabs.harvard.edu/abs/2016MNRAS.463..655K}
}

@ARTICLE{Romero2023,
       author = {{Romero}, Charles E. and {Gaspari}, Massimo and {Schellenberger}, Gerrit and {Bhandarkar}, Tanay and {Devlin}, Mark and {Dicker}, Simon R. and {Forman}, William and {Khatri}, Rishi and {Kraft}, Ralph and {Di Mascolo}, Luca and {Mason}, Brian S. and {Moravec}, Emily and {Mroczkowski}, Tony and {Nulsen}, Paul and {Orlowski-Scherer}, John and {Perez Sarmiento}, Karen and {Sarazin}, Craig and {Sievers}, Jonathan and {Su}, Yuanyuan},
        title = "{Inferences from Surface Brightness Fluctuations of Zwicky 3146 via the Sunyaev{\textendash}Zel'dovich Effect and X-Ray Observations}",
      journal = {\apj},
         year = 2023,
        month = jul,
       volume = {951},
       number = {1},
          eid = {41},
        pages = {41},
          doi = {10.3847/1538-4357/acd3f0},
archivePrefix = {arXiv},
       eprint = {2305.05790},
 primaryClass = {astro-ph.CO},
       adsurl = {https://ui.adsabs.harvard.edu/abs/2023ApJ...951...41R}
}

@ARTICLE{Adam2025,
       author = {{Adam}, R. and {Eynard-Machet}, T. and {Bartalucci}, I. and {Cherouvrier}, D. and {Clerc}, N. and {Di Mascolo}, L. and {Dupourqu{\'e}}, S. and {Ferrari}, C. and {Mac{\'\i}as-P{\'e}rez}, J. -F. and {Pointecouteau}, E. and {Pratt}, G.~W.},
        title = "{PITSZI: Probing intra-cluster medium turbulence with Sunyaev{\textendash}Zel'dovich imaging: Application to the triple merging cluster MACS J0717.5+3745}",
      journal = {\aap},
         year = 2025,
        month = feb,
       volume = {694},
          eid = {A182},
        pages = {A182},
          doi = {10.1051/0004-6361/202452342},
archivePrefix = {arXiv},
       eprint = {2409.14804},
 primaryClass = {astro-ph.CO},
       adsurl = {https://ui.adsabs.harvard.edu/abs/2025A&A...694A.182A}
}

@ARTICLE{Gaspari2014,
       author = {{Gaspari}, M. and {Churazov}, E. and {Nagai}, D. and {Lau}, E.~T. and {Zhuravleva}, I.},
        title = "{The relation between gas density and velocity power spectra in galaxy clusters: High-resolution hydrodynamic simulations and the role of conduction}",
      journal = {\aap},
         year = 2014,
        month = sep,
       volume = {569},
          eid = {A67},
        pages = {A67},
          doi = {10.1051/0004-6361/201424043},
archivePrefix = {arXiv},
       eprint = {1404.5302},
 primaryClass = {astro-ph.CO},
       adsurl = {https://ui.adsabs.harvard.edu/abs/2014A&A...569A..67G}
}

@ARTICLE{Zhuravleva2023,
       author = {{Zhuravleva}, Irina and {Chen}, Mandy C. and {Churazov}, Eugene and {Schekochihin}, Alexander A. and {Zhang}, Congyao and {Nagai}, Daisuke},
        title = "{Indirect measurements of gas velocities in galaxy clusters: effects of ellipticity and cluster dynamic state}",
      journal = {\mnras},
         year = 2023,
        month = apr,
       volume = {520},
       number = {4},
        pages = {5157-5172},
          doi = {10.1093/mnras/stad470},
archivePrefix = {arXiv},
       eprint = {2210.11544},
 primaryClass = {astro-ph.CO},
       adsurl = {https://ui.adsabs.harvard.edu/abs/2023MNRAS.520.5157Z}
}

@ARTICLE{Kolmogorov1941,
       author = {{Kolmogorov}, A.},
        title = "{The Local Structure of Turbulence in Incompressible Viscous Fluid for Very Large Reynolds' Numbers}",
      journal = {Akademiia Nauk SSSR Doklady},
         year = 1941,
        month = jan,
       volume = {30},
        pages = {301-305},
       adsurl = {https://ui.adsabs.harvard.edu/abs/1941DoSSR..30..301K}
}

@ARTICLE{Hitomi2016,
       author = {{Hitomi Collaboration}},
        title = "{The quiescent intracluster medium in the core of the Perseus cluster}",
      journal = {\nat},
         year = 2016,
        month = jul,
       volume = {535},
       number = {7610},
        pages = {117-121},
          doi = {10.1038/nature18627},
archivePrefix = {arXiv},
       eprint = {1607.04487},
 primaryClass = {astro-ph.GA},
       adsurl = {https://ui.adsabs.harvard.edu/abs/2016Natur.535..117H}
}

@ARTICLE{XRISM2025b,
       author = {{XRISM Collaboration}},
        title = "{XRISM Forecast for the Coma Cluster: Stormy, with a Steep Power Spectrum}",
      journal = {\apjl},
         year = 2025,
        month = may,
       volume = {985},
       number = {1},
          eid = {L20},
        pages = {L20},
          doi = {10.3847/2041-8213/add2f6},
archivePrefix = {arXiv},
       eprint = {2504.20928},
 primaryClass = {astro-ph.HE},
       adsurl = {https://ui.adsabs.harvard.edu/abs/2025ApJ...985L..20X}
}

@ARTICLE{XRISM2025a,
       author = {{XRISM Collaboration}},
        title = "{XRISM Reveals Low Nonthermal Pressure in the Core of the Hot, Relaxed Galaxy Cluster A2029}",
      journal = {\apjl},
         year = 2025,
        month = mar,
       volume = {982},
       number = {1},
          eid = {L5},
        pages = {L5},
          doi = {10.3847/2041-8213/ada7cd},
archivePrefix = {arXiv},
       eprint = {2501.05514},
 primaryClass = {astro-ph.HE},
       adsurl = {https://ui.adsabs.harvard.edu/abs/2025ApJ...982L...5X}
}

@ARTICLE{2009MNRAS.398.2049F,
       author = {{Feroz}, Farhan and {Hobson}, Michael P. and {Zwart}, Jonathan T.~L. and {Saunders}, Richard D.~E. and {Grainge}, Keith J.~B.},
        title = "{Bayesian modelling of clusters of galaxies from multifrequency-pointed Sunyaev-Zel'dovich observations}",
      journal = {\mnras},
         year = 2009,
        month = oct,
       volume = {398},
       number = {4},
        pages = {2049-2060},
          doi = {10.1111/j.1365-2966.2009.15247.x},
archivePrefix = {arXiv},
       eprint = {0811.1199},
 primaryClass = {astro-ph},
       adsurl = {https://ui.adsabs.harvard.edu/abs/2009MNRAS.398.2049F}
}

@ARTICLE{2019MNRAS.486.2116P,
       author = {{Perrott}, Yvette C. and {Javid}, Kamran and {Carvalho}, Pedro and {Elwood}, Patrick J. and {Hobson}, Michael P. and {Lasenby}, Anthony N. and {Olamaie}, Malak and {Saunders}, Richard D.~E.},
        title = "{Sunyaev-Zel'dovich profile fitting with joint AMI-Planck analysis}",
      journal = {\mnras},
         year = 2019,
        month = jun,
       volume = {486},
       number = {2},
        pages = {2116-2128},
          doi = {10.1093/mnras/stz826},
archivePrefix = {arXiv},
       eprint = {1901.09980},
 primaryClass = {astro-ph.CO},
       adsurl = {https://ui.adsabs.harvard.edu/abs/2019MNRAS.486.2116P}
}

@ARTICLE{2018MNRAS.481.3853O,
       author = {{Olamaie}, Malak and {Hobson}, Michael P. and {Feroz}, Farhan and {Grainge}, Keith J.~B. and {Lasenby}, Anthony and {Perrott}, Yvette C. and {Rumsey}, Clare and {Saunders}, Richard D.~E.},
        title = "{Free-form modelling of galaxy clusters: a Bayesian and data-driven approach}",
      journal = {\mnras},
         year = 2018,
        month = dec,
       volume = {481},
       number = {3},
        pages = {3853-3864},
          doi = {10.1093/mnras/sty2495},
archivePrefix = {arXiv},
       eprint = {1705.10712},
 primaryClass = {astro-ph.CO},
       adsurl = {https://ui.adsabs.harvard.edu/abs/2018MNRAS.481.3853O}
}

@ARTICLE{2025PASA...42...31G,
       author = {{Geris}, Sophia and {Perrott}, Yvette},
        title = "{Predicting Sunyaev-Zel'dovich effect observations of galaxy cluster cavities with the Square Kilometre Array}",
      journal = {\pasa},
         year = 2025,
        month = mar,
       volume = {42},
          eid = {e031},
        pages = {e031},
          doi = {10.1017/pasa.2025.13},
archivePrefix = {arXiv},
       eprint = {2503.23599},
 primaryClass = {astro-ph.CO},
       adsurl = {https://ui.adsabs.harvard.edu/abs/2025PASA...42...31G}
}

@ARTICLE{2011A&A...534L..12F,
       author = {{Ferrari}, C. and {Intema}, H.~T. and {Orr{\`u}}, E. and {Govoni}, F. and {Murgia}, M. and {Mason}, B. and {Bourdin}, H. and {Asad}, K.~M. and {Mazzotta}, P. and {Wise}, M.~W. and {Mroczkowski}, T. and {Croston}, J.~H.},
        title = "{Discovery of the correspondence between intra-cluster radio emission and a high pressure region detected through the Sunyaev-Zel'dovich effect}",
      journal = {\aap},
         year = 2011,
        month = oct,
       volume = {534},
          eid = {L12},
        pages = {L12},
          doi = {10.1051/0004-6361/201117788},
archivePrefix = {arXiv},
       eprint = {1107.5945},
 primaryClass = {astro-ph.CO},
       adsurl = {https://ui.adsabs.harvard.edu/abs/2011A&A...534L..12F}
}

@ARTICLE{2022A&A...667L...6O,
       author = {{Orlowski-Scherer}, John and {Haridas}, Saianeesh K. and {Di Mascolo}, Luca and {Sarmiento}, Karen Perez and {Romero}, Charles E. and {Dicker}, Simon and {Mroczkowski}, Tony and {Bhandarkar}, Tanay and {Churazov}, Eugene and {Clarke}, Tracy E. and {Devlin}, Mark and {Gaspari}, Massimo and {Lowe}, Ian and {Mason}, Brian and {Sarazin}, Craig L. and {Sievers}, Jonathon and {Sunyaev}, Rashid},
        title = "{GBT/MUSTANG-2 9″ resolution imaging of the SZ effect in MS0735.6+7421. Confirmation of the SZ cavities through direct imaging}",
      journal = {\aap},
         year = 2022,
        month = nov,
       volume = {667},
          eid = {L6},
        pages = {L6},
          doi = {10.1051/0004-6361/202244547},
archivePrefix = {arXiv},
       eprint = {2207.07100},
 primaryClass = {astro-ph.HE},
       adsurl = {https://ui.adsabs.harvard.edu/abs/2022A&A...667L...6O}
}

@ARTICLE{2018A&A...614A.118A,
       author = {{Adam}, R. and {Hahn}, O. and {Ruppin}, F. and {Ade}, P. and {Andr{\'e}}, P. and {Arnaud}, M. and {Bartalucci}, I. and {Beelen}, A. and {Beno{\^\i}t}, A. and {Bideaud}, A. and {Billot}, N. and {Bourrion}, O. and {Calvo}, M. and {Catalano}, A. and {Coiffard}, G. and {Comis}, B. and {D'Addabbo}, A. and {D{\'e}sert}, F.-X. and {Doyle}, S. and {Ferrari}, C. and {Goupy}, J. and {Kramer}, C. and {Lagache}, G. and {Leclercq}, S. and {Lestrade}, J.-F. and {Mac{\'\i}as-P{\'e}rez}, J.~F. and {Martinez Aviles}, G. and {Martizzi}, D. and {Maurogordato}, S. and {Mauskopf}, P. and {Mayet}, F. and {Monfardini}, A. and {Pajot}, F. and {Pascale}, E. and {Perotto}, L. and {Pisano}, G. and {Pointecouteau}, E. and {Ponthieu}, N. and {Pratt}, G.~W. and {Rev{\'e}ret}, V. and {Ricci}, M. and {Ritacco}, A. and {Rodriguez}, L. and {Romero}, C. and {Roussel}, H. and {Schuster}, K. and {Sievers}, A. and {Triqueneaux}, S. and {Tucker}, C. and {Wu}, H.-Y. and {Zylka}, R.},
        title = "{Substructure and merger detection in resolved NIKA Sunyaev-Zel'dovich images of distant clusters}",
      journal = {\aap},
         year = 2018,
        month = jun,
       volume = {614},
          eid = {A118},
        pages = {A118},
          doi = {10.1051/0004-6361/201731950},
archivePrefix = {arXiv},
       eprint = {1712.01836},
 primaryClass = {astro-ph.CO},
       adsurl = {https://ui.adsabs.harvard.edu/abs/2018A&A...614A.118A}
}

@ARTICLE{2018A&A...615A.112R,
       author = {{Ruppin}, F. and {Mayet}, F. and {Pratt}, G.~W. and {Adam}, R. and {Ade}, P. and {Andr{\'e}}, P. and {Arnaud}, M. and {Aussel}, H. and {Bartalucci}, I. and {Beelen}, A. and {Beno{\^\i}t}, A. and {Bideaud}, A. and {Bourrion}, O. and {Calvo}, M. and {Catalano}, A. and {Comis}, B. and {De Petris}, M. and {D{\'e}sert}, F.-X. and {Doyle}, S. and {Driessen}, E.~F.~C. and {Goupy}, J. and {Kramer}, C. and {Lagache}, G. and {Leclercq}, S. and {Lestrade}, J.-F. and {Mac{\'\i}as-P{\'e}rez}, J.~F. and {Mauskopf}, P. and {Monfardini}, A. and {Perotto}, L. and {Pisano}, G. and {Pointecouteau}, E. and {Ponthieu}, N. and {Rev{\'e}ret}, V. and {Ritacco}, A. and {Romero}, C. and {Roussel}, H. and {Schuster}, K. and {Sievers}, A. and {Tucker}, C. and {Zylka}, R.},
        title = "{First Sunyaev-Zel'dovich mapping with the NIKA2 camera: Implication of cluster substructures for the pressure profile and mass estimate}",
      journal = {\aap},
         year = 2018,
        month = jul,
       volume = {615},
          eid = {A112},
        pages = {A112},
          doi = {10.1051/0004-6361/201732558},
archivePrefix = {arXiv},
       eprint = {1712.09587},
 primaryClass = {astro-ph.CO},
       adsurl = {https://ui.adsabs.harvard.edu/abs/2018A&A...615A.112R}
}

@ARTICLE{2024A&A...684A..18A,
       author = {{Adam}, R. and {Ricci}, M. and {Eckert}, D. and {Ade}, P. and {Ajeddig}, H. and {Altieri}, B. and {Andr{\'e}}, P. and {Artis}, E. and {Aussel}, H. and {Beelen}, A. and {Benoist}, C. and {Beno{\^\i}t}, A. and {Berta}, S. and {Bing}, L. and {Birkinshaw}, M. and {Bourrion}, O. and {Boutigny}, D. and {Bremer}, M. and {Calvo}, M. and {Cappi}, A. and {Catalano}, A. and {De Petris}, M. and {D{\'e}sert}, F.-X. and {Doyle}, S. and {Driessen}, E.~F.~C. and {Faccioli}, L. and {Ferrari}, C. and {Gastaldello}, F. and {Giles}, P. and {Gomez}, A. and {Goupy}, J. and {Hahn}, O. and {Hanser}, C. and {Horellou}, C. and {K{\'e}ruzor{\'e}}, F. and {Koulouridis}, E. and {Kramer}, C. and {Ladjelate}, B. and {Lagache}, G. and {Leclercq}, S. and {Lestrade}, J.-F. and {Mac{\'\i}as-P{\'e}rez}, J.~F. and {Madden}, S. and {Maughan}, B. and {Maurogordato}, S. and {Maury}, A. and {Mauskopf}, P. and {Monfardini}, A. and {Mu{\~n}oz-Echeverr{\'\i}a}, M. and {Pacaud}, F. and {Perotto}, L. and {Pierre}, M. and {Pisano}, G. and {Pompei}, E. and {Ponthieu}, N. and {Rev{\'e}ret}, V. and {Rigby}, A. and {Ritacco}, A. and {Romero}, C. and {Roussel}, H. and {Ruppin}, F. and {Sereno}, M. and {Schuster}, K. and {Sievers}, A. and {Tintor{\'e} Vidal}, G. and {Tucker}, C. and {Zylka}, R.},
        title = "{The XXL Survey. LI. Pressure profile and Y$_{SZ}$ {\ensuremath{-}} M scaling relation in three low-mass galaxy clusters at z {\ensuremath{\sim}} 1 observed with NIKA2}",
      journal = {\aap},
         year = 2024,
        month = apr,
       volume = {684},
          eid = {A18},
        pages = {A18},
          doi = {10.1051/0004-6361/202348049},
archivePrefix = {arXiv},
       eprint = {2310.05819},
 primaryClass = {astro-ph.CO},
       adsurl = {https://ui.adsabs.harvard.edu/abs/2024A&A...684A..18A}
}

@ARTICLE{2025A&A...702A.275M,
       author = {{Mu{\~n}oz-Echeverr{\'\i}a}, M. and {Mac{\'\i}as-P{\'e}rez}, J.-F. and {Neri}, R. and {Pointecouteau}, E. and {Adam}, R. and {Ade}, P. and {Ajeddig}, H. and {Amarantidis}, S. and {Andr{\'e}}, P. and {Aussel}, H. and {Beelen}, A. and {Beno{\^\i}t}, A. and {Berta}, S. and {B{\'e}thermin}, M. and {Bongiovanni}, A. and {Bounmy}, J. and {Bourrion}, O. and {Calvo}, M. and {Catalano}, A. and {Ch{\'e}rouvrier}, D. and {Chowdhury}, U. and {De Petris}, M. and {D{\'e}sert}, F.-X. and {Doyle}, S. and {Driessen}, E.~F.~C. and {Ejlali}, G. and {Ferragamo}, A. and {G{\'o}mez}, A. and {Goupy}, J. and {Hanser}, C. and {Katsioli}, S. and {K{\'e}ruzor{\'e}}, F. and {Kramer}, C. and {Ladjelate}, B. and {Lagache}, G. and {Leclercq}, S. and {Lestrade}, J.-F. and {Madden}, S.~C. and {Maury}, A. and {Mayet}, F. and {Monfardini}, A. and {Moyer-Anin}, A. and {Myserlis}, I. and {Paliwal}, A. and {Perotto}, L. and {Pisano}, G. and {Ponthieu}, N. and {Rev{\'e}ret}, V. and {Rigby}, A.~J. and {Ritacco}, A. and {Roussel}, H. and {Ruppin}, F. and {S{\'a}nchez-Portal}, M. and {Savorgnano}, S. and {Schuster}, K. and {Sievers}, A. and {Tucker}, C. and {Zylka}, R. and {Bartalucci}, I. and {Melin}, J.-B. and {Pratt}, G.~W.},
        title = "{Thermal Sunyaev-Zel'dovich effect at the core of CL J1226.9+3332 revealed by NOEMA}",
      journal = {\aap},
         year = 2025,
        month = oct,
       volume = {702},
          eid = {A275},
        pages = {A275},
          doi = {10.1051/0004-6361/202555851},
archivePrefix = {arXiv},
       eprint = {2509.14048},
 primaryClass = {astro-ph.CO},
       adsurl = {https://ui.adsabs.harvard.edu/abs/2025A&A...702A.275M}
}

@ARTICLE{2008SSRv..134...93F,
       author = {{Ferrari}, C. and {Govoni}, F. and {Schindler}, S. and {Bykov}, A.~M. and {Rephaeli}, Y.},
        title = "{Observations of Extended Radio Emission in Clusters}",
      journal = {\ssr},
         year = 2008,
        month = feb,
       volume = {134},
       number = {1-4},
        pages = {93-118},
          doi = {10.1007/s11214-008-9311-x},
archivePrefix = {arXiv},
       eprint = {0801.0985},
 primaryClass = {astro-ph},
       adsurl = {https://ui.adsabs.harvard.edu/abs/2008SSRv..134...93F}
}

@ARTICLE{1970Ap&SS...7....3S,
       author = {{Sunyaev}, R.~A. and {Zeldovich}, Ya. B.},
        title = "{Small-Scale Fluctuations of Relic Radiation}",
      journal = {\apss},
         year = 1970,
        month = apr,
       volume = {7},
       number = {1},
        pages = {3-19},
          doi = {10.1007/BF00653471},
       adsurl = {https://ui.adsabs.harvard.edu/abs/1970Ap&SS...7....3S}
}

@ARTICLE{2019SSRv..215...17M,
       author = {{Mroczkowski}, Tony and {Nagai}, Daisuke and {Basu}, Kaustuv and {Chluba}, Jens and {Sayers}, Jack and {Adam}, R{\'e}mi and {Churazov}, Eugene and {Crites}, Abigail and {Di Mascolo}, Luca and {Eckert}, Dominique and {Macias-Perez}, Juan and {Mayet}, Fr{\'e}d{\'e}ric and {Perotto}, Laurence and {Pointecouteau}, Etienne and {Romero}, Charles and {Ruppin}, Florian and {Scannapieco}, Evan and {ZuHone}, John},
        title = "{Astrophysics with the Spatially and Spectrally Resolved Sunyaev-Zeldovich Effects. A Millimetre/Submillimetre Probe of the Warm and Hot Universe}",
      journal = {\ssr},
         year = 2019,
        month = feb,
       volume = {215},
       number = {1},
          eid = {17},
        pages = {17},
          doi = {10.1007/s11214-019-0581-2},
archivePrefix = {arXiv},
       eprint = {1811.02310},
 primaryClass = {astro-ph.CO},
       adsurl = {https://ui.adsabs.harvard.edu/abs/2019SSRv..215...17M}
}

@ARTICLE{2024A&A...689A..41V,
       author = {{van Marrewijk}, J. and {Di Mascolo}, L. and {Gill}, A.~S. and {Battaglia}, N. and {Battistelli}, E.~S. and {Bond}, J.~R. and {Devlin}, M.~J. and {Doze}, P. and {Dunkley}, J. and {Knowles}, K. and {Hincks}, A. and {Hughes}, J.~P. and {Hilton}, M. and {Moodley}, K. and {Mroczkowski}, T. and {Naess}, S. and {Partridge}, B. and {Popping}, G. and {Sif{\'o}n}, C. and {Staggs}, S.~T. and {Wollack}, E.~J.},
        title = "{XLSSC 122 caught in the act of growing up: Spatially resolved SZ observations of a z = 1.98 galaxy cluster}",
      journal = {\aap},
         year = 2024,
        month = sep,
       volume = {689},
          eid = {A41},
        pages = {A41},
          doi = {10.1051/0004-6361/202348213},
archivePrefix = {arXiv},
       eprint = {2310.06120},
 primaryClass = {astro-ph.CO},
       adsurl = {https://ui.adsabs.harvard.edu/abs/2024A&A...689A..41V}
}

@ARTICLE{2023Natur.615..809D,
       author = {{Di Mascolo}, Luca and {Saro}, Alexandro and {Mroczkowski}, Tony and {Borgani}, Stefano and {Churazov}, Eugene and {Rasia}, Elena and {Tozzi}, Paolo and {Dannerbauer}, Helmut and {Basu}, Kaustuv and {Carilli}, Christopher L. and {Ginolfi}, Michele and {Miley}, George and {Nonino}, Mario and {Pannella}, Maurilio and {Pentericci}, Laura and {Rizzo}, Francesca},
        title = "{Forming intracluster gas in a galaxy protocluster at a redshift of 2.16}",
      journal = {\nat},
         year = 2023,
        month = mar,
       volume = {615},
       number = {7954},
        pages = {809-812},
          doi = {10.1038/s41586-023-05761-x},
archivePrefix = {arXiv},
       eprint = {2303.16226},
 primaryClass = {astro-ph.CO},
       adsurl = {https://ui.adsabs.harvard.edu/abs/2023Natur.615..809D}
}

@ARTICLE{2019A&A...621A..41G,
       author = {{Ghirardini}, V. and {Eckert}, D. and {Ettori}, S. and {Pointecouteau}, E. and {Molendi}, S. and {Gaspari}, M. and {Rossetti}, M. and {De Grandi}, S. and {Roncarelli}, M. and {Bourdin}, H. and {Mazzotta}, P. and {Rasia}, E. and {Vazza}, F.},
        title = "{Universal thermodynamic properties of the intracluster medium over two decades in radius in the X-COP sample}",
      journal = {\aap},
         year = 2019,
        month = jan,
       volume = {621},
          eid = {A41},
        pages = {A41},
          doi = {10.1051/0004-6361/201833325},
archivePrefix = {arXiv},
       eprint = {1805.00042},
 primaryClass = {astro-ph.CO},
       adsurl = {https://ui.adsabs.harvard.edu/abs/2019A&A...621A..41G}
}

@ARTICLE{2016ApJ...829L..23B,
       author = {{Basu}, K. and {Sommer}, M. and {Erler}, J. and {Eckert}, D. and {Vazza}, F. and {Magnelli}, B. and {Bertoldi}, F. and {Tozzi}, P.},
        title = "{ALMA-SZ Detection of a Galaxy Cluster Merger Shock at Half the Age of the Universe}",
      journal = {\apjl},
         year = 2016,
        month = oct,
       volume = {829},
       number = {2},
          eid = {L23},
        pages = {L23},
          doi = {10.3847/2041-8205/829/2/L23},
archivePrefix = {arXiv},
       eprint = {1608.05413},
 primaryClass = {astro-ph.CO},
       adsurl = {https://ui.adsabs.harvard.edu/abs/2016ApJ...829L..23B}
}

@ARTICLE{2011ARA&A..49..409A,
       author = {{Allen}, Steven W. and {Evrard}, August E. and {Mantz}, Adam B.},
        title = "{Cosmological Parameters from Observations of Galaxy Clusters}",
      journal = {\araa},
         year = 2011,
        month = sep,
       volume = {49},
       number = {1},
        pages = {409-470},
          doi = {10.1146/annurev-astro-081710-102514},
archivePrefix = {arXiv},
       eprint = {1103.4829},
 primaryClass = {astro-ph.CO},
       adsurl = {https://ui.adsabs.harvard.edu/abs/2011ARA&A..49..409A}
}

@ARTICLE{2012ARA&A..50..353K,
       author = {{Kravtsov}, Andrey V. and {Borgani}, Stefano},
        title = "{Formation of Galaxy Clusters}",
      journal = {\araa},
         year = 2012,
        month = sep,
       volume = {50},
        pages = {353-409},
          doi = {10.1146/annurev-astro-081811-125502},
archivePrefix = {arXiv},
       eprint = {1205.5556},
 primaryClass = {astro-ph.CO},
       adsurl = {https://ui.adsabs.harvard.edu/abs/2012ARA&A..50..353K}
}

@ARTICLE{2019SSRv..215...25P,
       author = {{Pratt}, G.~W. and {Arnaud}, M. and {Biviano}, A. and {Eckert}, D. and {Ettori}, S. and {Nagai}, D. and {Okabe}, N. and {Reiprich}, T.~H.},
        title = "{The Galaxy Cluster Mass Scale and Its Impact on Cosmological Constraints from the Cluster Population}",
      journal = {\ssr},
         year = 2019,
        month = feb,
       volume = {215},
       number = {2},
          eid = {25},
        pages = {25},
          doi = {10.1007/s11214-019-0591-0},
archivePrefix = {arXiv},
       eprint = {1902.10837},
 primaryClass = {astro-ph.CO},
       adsurl = {https://ui.adsabs.harvard.edu/abs/2019SSRv..215...25P}
}

@ARTICLE{SO2019,
       author = {{SO Collaboration}},
        title = "{The Simons Observatory: science goals and forecasts}",
      journal = {\jcap},
         year = 2019,
        month = feb,
       volume = {2019},
       number = {2},
          eid = {056},
        pages = {056},
          doi = {10.1088/1475-7516/2019/02/056},
archivePrefix = {arXiv},
       eprint = {1808.07445},
 primaryClass = {astro-ph.CO},
       adsurl = {https://ui.adsabs.harvard.edu/abs/2019JCAP...02..056A}
}

@ARTICLE{SOLAT2025,
       author = {{SO Collaboration}},
        title = "{The Simons Observatory: science goals and forecasts for the enhanced Large Aperture Telescope}",
      journal = {\jcap},
         year = 2025,
        month = aug,
       volume = {2025},
       number = {8},
          eid = {034},
        pages = {034},
          doi = {10.1088/1475-7516/2025/08/034},
archivePrefix = {arXiv},
       eprint = {2503.00636},
 primaryClass = {astro-ph.IM},
       adsurl = {https://ui.adsabs.harvard.edu/abs/2025JCAP...08..034A}
}

@ARTICLE{newathena25,
       author = {{Cruise}, Mike and {Guainazzi}, Matteo and {Aird}, James and {Carrera}, Francisco J. and {Costantini}, Elisa and {Corrales}, Lia and {Dauser}, Thomas and {Eckert}, Dominique and {Gastaldello}, Fabio and {Matsumoto}, Hironori and {Osten}, Rachel and {Petrucci}, Pierre-Olivier and {Porquet}, Delphine and {Pratt}, Gabriel W. and {Rea}, Nanda and {Reiprich}, Thomas H. and {Simionescu}, Aurora and {Spiga}, Daniele and {Troja}, Eleonora},
        title = "{The NewAthena mission concept in the context of the next decade of X-ray astronomy}",
      journal = {Nature Astronomy},
         year = 2025,
        month = jan,
       volume = {9},
        pages = {36-44},
          doi = {10.1038/s41550-024-02416-3},
archivePrefix = {arXiv},
       eprint = {2501.03100},
 primaryClass = {astro-ph.IM},
       adsurl = {https://ui.adsabs.harvard.edu/abs/2025NatAs...9...36C}
}

@ARTICLE{Markevitch2007,
       author = {{Markevitch}, Maxim and {Vikhlinin}, Alexey},
        title = "{Shocks and cold fronts in galaxy clusters}",
      journal = {\physrep},
         year = 2007,
        month = may,
       volume = {443},
       number = {1},
        pages = {1-53},
          doi = {10.1016/j.physrep.2007.01.001},
archivePrefix = {arXiv},
       eprint = {astro-ph/0701821},
 primaryClass = {astro-ph},
       adsurl = {https://ui.adsabs.harvard.edu/abs/2007PhR...443....1M}
}

@ARTICLE{Vink2015,
       author = {{Vink}, Jacco and {Broersen}, Sjors and {Bykov}, Andrei and {Gabici}, Stefano},
        title = "{On the electron-ion temperature ratio established by collisionless shocks}",
      journal = {\aap},
         year = 2015,
        month = jul,
       volume = {579},
          eid = {A13},
        pages = {A13},
          doi = {10.1051/0004-6361/201424612},
archivePrefix = {arXiv},
       eprint = {1407.4499},
 primaryClass = {astro-ph.HE},
       adsurl = {https://ui.adsabs.harvard.edu/abs/2015A&A...579A..13V}
}

@INPROCEEDINGS{Markevitch2006,
       author = {{Markevitch}, M.},
        title = "{Chandra Observation of the Most Interesting Cluster in the Universe}",
    booktitle = {The X-ray Universe 2005},
         year = 2006,
       editor = {{Wilson}, A.},
       series = {ESA Special Publication},
       volume = {604},
        month = jan,
        pages = {723},
          doi = {10.48550/arXiv.astro-ph/0511345},
archivePrefix = {arXiv},
       eprint = {astro-ph/0511345},
 primaryClass = {astro-ph},
       adsurl = {https://ui.adsabs.harvard.edu/abs/2006ESASP.604..723M}
}

@ARTICLE{Russell2022,
       author = {{Russell}, H.~R. and {Nulsen}, P.~E.~J. and {Caprioli}, D. and {Chadayammuri}, U. and {Fabian}, A.~C. and {Kunz}, M.~W. and {McNamara}, B.~R. and {Sanders}, J.~S. and {Richard-Laferri{\`e}re}, A. and {Beleznay}, M. and {Canning}, R.~E.~A. and {Hlavacek-Larrondo}, J. and {King}, L.~J.},
        title = "{The structure of cluster merger shocks: turbulent width and the electron heating time-scale}",
      journal = {\mnras},
         year = 2022,
        month = jul,
       volume = {514},
       number = {1},
        pages = {1477-1493},
          doi = {10.1093/mnras/stac1055},
archivePrefix = {arXiv},
       eprint = {2204.05785},
 primaryClass = {astro-ph.HE},
       adsurl = {https://ui.adsabs.harvard.edu/abs/2022MNRAS.514.1477R}
}

@ARTICLE{Bourdin2013,
       author = {{Bourdin}, H. and {Mazzotta}, P. and {Markevitch}, M. and {Giacintucci}, S. and {Brunetti}, G.},
        title = "{Shock Heating of the Merging Galaxy Cluster A521}",
      journal = {\apj},
         year = 2013,
        month = feb,
       volume = {764},
       number = {1},
          eid = {82},
        pages = {82},
          doi = {10.1088/0004-637X/764/1/82},
archivePrefix = {arXiv},
       eprint = {1302.0696},
 primaryClass = {astro-ph.CO},
       adsurl = {https://ui.adsabs.harvard.edu/abs/2013ApJ...764...82B}
}

@ARTICLE{Norseth2025,
       author = {{Norseth}, Christian T. and {Wik}, Daniel R. and {Sarazin}, Craig L. and {Sun}, Ming and {Gastaldello}, Fabio},
        title = "{Electron{\textendash}Ion Equilibration in the Merging Galaxy Cluster A665}",
      journal = {\apj},
         year = 2025,
        month = oct,
       volume = {992},
       number = {1},
          eid = {62},
        pages = {62},
          doi = {10.3847/1538-4357/adfe6b},
archivePrefix = {arXiv},
       eprint = {2508.15138},
 primaryClass = {astro-ph.CO},
       adsurl = {https://ui.adsabs.harvard.edu/abs/2025ApJ...992...62N}
}

@ARTICLE{Wang2016,
       author = {{Wang}, Qian H.~S. and {Markevitch}, Maxim and {Giacintucci}, Simona},
        title = "{The Merging Galaxy Cluster A520{\textemdash}A Broken-up Cool Core, A Dark Subcluster, and an X-Ray Channel}",
      journal = {\apj},
         year = 2016,
        month = dec,
       volume = {833},
       number = {1},
          eid = {99},
        pages = {99},
          doi = {10.3847/1538-4357/833/1/99},
archivePrefix = {arXiv},
       eprint = {1603.05232},
 primaryClass = {astro-ph.CO},
       adsurl = {https://ui.adsabs.harvard.edu/abs/2016ApJ...833...99W}
}

@ARTICLE{Wang2018,
       author = {{Wang}, Qian H.~S. and {Giacintucci}, Simona and {Markevitch}, Maxim},
        title = "{Bow Shock in Merging Cluster A520: The Edge of the Radio Halo and the Electron-Proton Equilibration Timescale}",
      journal = {\apj},
         year = 2018,
        month = apr,
       volume = {856},
       number = {2},
          eid = {162},
        pages = {162},
          doi = {10.3847/1538-4357/aab2aa},
archivePrefix = {arXiv},
       eprint = {1802.03402},
 primaryClass = {astro-ph.HE},
       adsurl = {https://ui.adsabs.harvard.edu/abs/2018ApJ...856..162W}
}

@ARTICLE{2019A&A...628A.100D,
       author = {{Di Mascolo}, Luca and {Mroczkowski}, Tony and {Churazov}, Eugene and {Markevitch}, Maxim and {Basu}, Kaustuv and {Clarke}, Tracy E. and {Devlin}, Mark and {Mason}, Brian S. and {Randall}, Scott W. and {Reese}, Erik D. and {Sunyaev}, Rashid and {Wik}, Daniel R.},
        title = "{An ALMA+ACA measurement of the shock in the Bullet Cluster}",
      journal = {\aap},
         year = 2019,
        month = aug,
       volume = {628},
          eid = {A100},
        pages = {A100},
          doi = {10.1051/0004-6361/201936184},
archivePrefix = {arXiv},
       eprint = {1907.07680},
 primaryClass = {astro-ph.CO},
       adsurl = {https://ui.adsabs.harvard.edu/abs/2019A&A...628A.100D}
}

@ARTICLE{2021A&A...651A..41C,
       author = {{Churazov}, E. and {Khabibullin}, I. and {Lyskova}, N. and {Sunyaev}, R. and {Bykov}, A.~M.},
        title = "{Tempestuous life beyond R$_{500}$: X-ray view on the Coma cluster with SRG/eROSITA. I. X-ray morphology, recent merger, and radio halo connection}",
      journal = {\aap},
         year = 2021,
        month = jul,
       volume = {651},
          eid = {A41},
        pages = {A41},
          doi = {10.1051/0004-6361/202040197},
archivePrefix = {arXiv},
       eprint = {2012.11627},
 primaryClass = {astro-ph.HE},
       adsurl = {https://ui.adsabs.harvard.edu/abs/2021A&A...651A..41C}
}

@ARTICLE{2015ApJ...813..129Z,
       author = {{Zhang}, Congyao and {Yu}, Qingjuan and {Lu}, Youjun},
        title = "{Simulating the Galaxy Cluster {\textquotedblleft}El Gordo{\textquotedblright} and Identifying the Merger Configuration}",
      journal = {\apj},
         year = 2015,
        month = nov,
       volume = {813},
       number = {2},
          eid = {129},
        pages = {129},
          doi = {10.1088/0004-637X/813/2/129},
archivePrefix = {arXiv},
       eprint = {1511.02578},
 primaryClass = {astro-ph.CO},
       adsurl = {https://ui.adsabs.harvard.edu/abs/2015ApJ...813..129Z}
}

@ARTICLE{2018ApJ...855...36Z,
       author = {{Zhang}, Congyao and {Yu}, Qingjuan and {Lu}, Youjun},
        title = "{Simulating the Galaxy Cluster {\textquotedblleft}El Gordo{\textquotedblright}: Gas Motion, Kinetic Sunyaev-Zel{\textquoteright}dovich Signal, and X-Ray Line Features}",
      journal = {\apj},
         year = 2018,
        month = mar,
       volume = {855},
       number = {1},
          eid = {36},
        pages = {36},
          doi = {10.3847/1538-4357/aaab4c},
archivePrefix = {arXiv},
       eprint = {1711.08438},
 primaryClass = {astro-ph.CO},
       adsurl = {https://ui.adsabs.harvard.edu/abs/2018ApJ...855...36Z}
}

@ARTICLE{Marriage2011,
       author = {{Marriage}, Tobias A. and {Acquaviva}, Viviana and {Ade}, Peter A.~R. and {Aguirre}, Paula and {Amiri}, Mandana and {Appel}, John William and {Barrientos}, L. Felipe and {Battistelli}, Elia S. and {Bond}, J. Richard and {Brown}, Ben and {Burger}, Bryce and {Chervenak}, Jay and {Das}, Sudeep and {Devlin}, Mark J. and {Dicker}, Simon R. and {Bertrand Doriese}, W. and {Dunkley}, Joanna and {D{\"u}nner}, Rolando and {Essinger-Hileman}, Thomas and {Fisher}, Ryan P. and {Fowler}, Joseph W. and {Hajian}, Amir and {Halpern}, Mark and {Hasselfield}, Matthew and {Hern{\'a}ndez-Monteagudo}, Carlos and {Hilton}, Gene C. and {Hilton}, Matt and {Hincks}, Adam D. and {Hlozek}, Ren{\'e}e and {Huffenberger}, Kevin M. and {Handel Hughes}, David and {Hughes}, John P. and {Infante}, Leopoldo and {Irwin}, Kent D. and {Baptiste Juin}, Jean and {Kaul}, Madhuri and {Klein}, Jeff and {Kosowsky}, Arthur and {Lau}, Judy M. and {Limon}, Michele and {Lin}, Yen-Ting and {Lupton}, Robert H. and {Marsden}, Danica and {Martocci}, Krista and {Mauskopf}, Phil and {Menanteau}, Felipe and {Moodley}, Kavilan and {Moseley}, Harvey and {Netterfield}, Calvin B. and {Niemack}, Michael D. and {Nolta}, Michael R. and {Page}, Lyman A. and {Parker}, Lucas and {Partridge}, Bruce and {Quintana}, Hernan and {Reese}, Erik D. and {Reid}, Beth and {Sehgal}, Neelima and {Sherwin}, Blake D. and {Sievers}, Jon and {Spergel}, David N. and {Staggs}, Suzanne T. and {Swetz}, Daniel S. and {Switzer}, Eric R. and {Thornton}, Robert and {Trac}, Hy and {Tucker}, Carole and {Warne}, Ryan and {Wilson}, Grant and {Wollack}, Ed and {Zhao}, Yue},
        title = "{The Atacama Cosmology Telescope: Sunyaev-Zel'dovich-Selected Galaxy Clusters at 148 GHz in the 2008 Survey}",
      journal = {\apj},
         year = 2011,
        month = aug,
       volume = {737},
       number = {2},
          eid = {61},
        pages = {61},
          doi = {10.1088/0004-637X/737/2/61},
archivePrefix = {arXiv},
       eprint = {1010.1065},
 primaryClass = {astro-ph.CO},
       adsurl = {https://ui.adsabs.harvard.edu/abs/2011ApJ...737...61M}
}

@ARTICLE{AXIS2025,
       author = {{Koss}, Michael and {Aftab}, Nafisa and {Allen}, Steven W. and {Amato}, Roberta and {An}, Hongjun and {Andreoni}, Igor and {Anguita}, Timo and {Arcodia}, Riccardo and {Ayres}, Thomas and {Bachetti}, Matteo and {Baglio}, Maria Cristina and {Bahramian}, Arash and {Balboni}, Marco and {Baldi}, Ranieri D. and {Balman}, Solen and {Bamba}, Aya and {Banados}, Eduardo and {Bao}, Tong and {Bartalucci}, Iacopo and {Basu-Zych}, Antara and {Batalha}, Rebeca and {Battistini}, Lorenzo and {Bauer}, Franz Erik and {Beardmore}, Andy and {Becker}, Werner and {Behar}, Ehud and {Belfiore}, Andrea and {Beniamini}, Paz and {Bertola}, Elena and {Bessa}, Vinicius and {Best}, Henry and {Bianchi}, Stefano and {Biava}, N. and {Binder}, Breanna A. and {Blanton}, Elizabeth L. and {Bodaghee}, Arash and {Bogdanovic}, Tamara and {Bogensberger}, David and {Bonafede}, A. and {Bonetti}, Matteo and {Bordas}, Pol and {Borghese}, Alice and {Botteon}, Andrea and {Boula}, Stella and {Bozzo}, Enrico and {Branchesi}, Marica and {Brandt}, William Nielsen and {Bregman}, Joel and {Brighenti}, Fabrizio and {Bronzini}, Ettore and {Brunelli}, Giulia and {Brusa}, Marcella and {Bulbul}, Esra and {Burdge}, Kevin and {Caccianiga}, Alessandro and {Calzadilla}, Michael and {Campana}, Sergio and {Capalbi}, Milvia and {Capitanio}, Fiamma and {Cappelluti}, Nico and {Carney}, Jonathan and {Casanova}, Sabrina and {Castro}, Daniel and {Cenko}, S. Bradley and {Chakraborty}, Joheen and {Chakraborty}, Priyanka and {Chartas}, George and {Chatterjee}, Arka and {Choudhury}, Prakriti Pal and {Cilley}, Raven and {Civano}, Francesca and {Comastri}, Andrea and {Connor}, Thomas and {Corcoran}, Michael F. and {Corrales}, Lia and {Coti Zelati}, Francesco and {Cui}, Weiguang and {D'Ammando}, Filippo and {Dage}, Kristen and {Daylan}, Tansu and {De Grandi}, Sabrina and {De Rosa}, Alessandra and {Decarli}, Roberto and {Decourchelle}, Anne and {Degenaar}, Nathalie and {Del Popolo}, Antonino and {Di Marco}, Alessandro and {Di Salvo}, Tiziana and {Dichiara}, Simone and {DiKerby}, Stephen and {Dillmann}, Steven and {Doerksen}, Neil and {Draghis}, Paul and {Drake}, Jeremy J. and {Ducci}, Lorenzo and {Dupke}, Renato and {Durbak}, Joseph and {Duvvuri}, Girish M. and {Dykaar}, Hannah and {Eckert}, Dominique and {Elvis}, Martin and {Espaillat}, Catherine and {Esposito}, Paolo and {Furst}, Felix and {Fabbiano}, Giuseppina and {Fagin}, Joshua and {Falcone}, Abraham and {Fedorova}, Elena and {Feinstein}, Adina and {Fernandez Fernandez}, Jorge and {Ferrand}, Gilles and {Flores}, Anthony M. and {Foo}, N. and {Foo}, Nicholas and {Foord}, Adi and {Franchini}, Alessia and {Fraschetti}, Federico and {Frye}, Brenda L. and {Lowenthal}, James D. and {Fryer}, Chris and {Fujimoto}, Shin-ichiro and {Gagnon}, Seth and {Gallo}, Luigi and {Garcia Diaz}, Carlos and {Gaspari}, Massimo and {Gastaldello}, Fabio and {Gelfand}, Joseph D. and {Gezari}, Suvi and {Ghizzardi}, Simona and {Giacintucci}, Simona and {Gill}, A. and {Gilli}, Roberto and {Gitti}, Myriam and {Giustini}, Margherita and {Gnarini}, Andrea and {Grandi}, Paola and {Gross}, Arran and {Gu}, Liyi and {Gunderson}, Sean and {Gunther}, Hans Moritz and {Haggard}, Daryl and {Hamaguchi}, Kenji and {Hare}, Jeremy and {Harrington}, Kevin C. and {Heinke}, Craig and {Heinz}, Sebastian and {Hlavacek-Larrondo}, Julie and {Ho}, Wynn C.~G. and {Hodges-Kluck}, Edmund and {Homan}, Jeroen and {Huang}, R. and {Ighina}, Luca and {Ignesti}, Alessandro and {Imbrogno}, Matteo and {Irwin}, Christopher and {Irwin}, Jimmy and {Islam}, Nazma and {Israel}, Gian Luca and {Jacobson-Galan}, Wynn and {Jain}, Chetana and {Jana}, Arghajit and {Jaodand}, Amruta and {Jennings}, Fred and {Jiang}, Jiachen and {Jimenez-Andrade}, Eric F. and {Jimenez-Teja}, Y. and {Johnson}, S. and {Jonker}, Peter and {Kamieneski}, Patrick S. and {Kammoun}, Elias and {Kara}, Erin and {Kargaltsev}, Oleg and {King}, George W. and {Kirmizibayrak}, Demet and {Klingler}, Noel and {Kong}, Albert K.~H. and {Kounkel}, Marina and {Kumar}, Manish and {Kutyrev}, Alexander and {Kyer}, Rebecca and {La Monaca}, Fabio and {Lambrides}, Erini and {Lanzuisi}, Giorgio and {Lee}, Wonki and {Lehmer}, Bret and {Lentini}, Elisa and {Lepore}, Marika and {Li}, Jiangtao and {Lisse}, Carey M. and {Liu}, Daizhong and {Liu}, Tingting and {Isla Llave}, Monica and {Locatelli}, Nicola and {Lopez}, Laura A. and {Lopez}, Sebastian and {Lovisari}, Lorenzo and {Lusso}, Elisabeta and {Mac Intyre}, Brydyn and {MacMaster}, Austin and {Maiolino}, Roberto},
        title = "{The Advanced X-ray Imaging Satellite Community Science Book}",
      journal = {arXiv e-prints},
         year = 2025,
        month = oct,
          eid = {arXiv:2511.00253},
        pages = {arXiv:2511.00253},
archivePrefix = {arXiv},
       eprint = {2511.00253},
 primaryClass = {astro-ph.HE},
       adsurl = {https://ui.adsabs.harvard.edu/abs/2025arXiv251100253K}
}

@ARTICLE{2019ApJ...871..195A,
       author = {{Abdulla}, Zubair and {Carlstrom}, John E. and {Mantz}, Adam B. and {Marrone}, Daniel P. and {Greer}, Christopher H. and {Lamb}, James W. and {Leitch}, Erik M. and {Muchovej}, Stephen and {O'Donnell}, Christine and {Plagge}, Thomas J. and {Woody}, David},
        title = "{Constraints on the Thermal Contents of the X-Ray Cavities of Cluster MS 0735.6+7421 with Sunyaev-Zel{\textquoteright}dovich Effect Observations}",
      journal = {\apj},
         year = 2019,
        month = feb,
       volume = {871},
       number = {2},
          eid = {195},
        pages = {195},
          doi = {10.3847/1538-4357/aaf888},
archivePrefix = {arXiv},
       eprint = {1806.05050},
 primaryClass = {astro-ph.GA},
       adsurl = {https://ui.adsabs.harvard.edu/abs/2019ApJ...871..195A}
}

@ARTICLE{2014JLTP..176..808D,
       author = {{Dicker}, S.~R. and {Ade}, P.~A.~R. and {Aguirre}, J. and {Brevik}, J.~A. and {Cho}, H.~M. and {Datta}, R. and {Devlin}, M.~J. and {Dober}, B. and {Egan}, D. and {Ford}, J. and {Ford}, P. and {Hilton}, G. and {Irwin}, K.~D. and {Mason}, B.~S. and {Marganian}, P. and {Mello}, M. and {McMahon}, J.~J. and {Mroczkowski}, T. and {Rosenman}, M. and {Tucker}, C. and {Vale}, L. and {White}, S. and {Whitehead}, M. and {Young}, A.~H.},
        title = "{MUSTANG 2: A Large Focal Plane Array for the 100 m Green Bank Telescope}",
      journal = {Journal of Low Temperature Physics},
         year = 2014,
        month = sep,
       volume = {176},
       number = {5-6},
        pages = {808-814},
          doi = {10.1007/s10909-013-1070-8},
       adsurl = {https://ui.adsabs.harvard.edu/abs/2014JLTP..176..808D}
}

@ARTICLE{2018A&A...609A.115A,
       author = {{Adam}, R. and {Adane}, A. and {Ade}, P.~A.~R. and {Andr{\'e}}, P. and {Andrianasolo}, A. and {Aussel}, H. and {Beelen}, A. and {Beno{\^\i}t}, A. and {Bideaud}, A. and {Billot}, N. and {Bourrion}, O. and {Bracco}, A. and {Calvo}, M. and {Catalano}, A. and {Coiffard}, G. and {Comis}, B. and {De Petris}, M. and {D{\'e}sert}, F.-X. and {Doyle}, S. and {Driessen}, E.~F.~C. and {Evans}, R. and {Goupy}, J. and {Kramer}, C. and {Lagache}, G. and {Leclercq}, S. and {Leggeri}, J.-P. and {Lestrade}, J.-F. and {Mac{\'\i}as-P{\'e}rez}, J.~F. and {Mauskopf}, P. and {Mayet}, F. and {Maury}, A. and {Monfardini}, A. and {Navarro}, S. and {Pascale}, E. and {Perotto}, L. and {Pisano}, G. and {Ponthieu}, N. and {Rev{\'e}ret}, V. and {Rigby}, A. and {Ritacco}, A. and {Romero}, C. and {Roussel}, H. and {Ruppin}, F. and {Schuster}, K. and {Sievers}, A. and {Triqueneaux}, S. and {Tucker}, C. and {Zylka}, R.},
        title = "{The NIKA2 large-field-of-view millimetre continuum camera for the 30 m IRAM telescope}",
      journal = {\aap},
         year = 2018,
        month = jan,
       volume = {609},
          eid = {A115},
        pages = {A115},
          doi = {10.1051/0004-6361/201731503},
archivePrefix = {arXiv},
       eprint = {1707.00908},
 primaryClass = {astro-ph.IM},
       adsurl = {https://ui.adsabs.harvard.edu/abs/2018A&A...609A.115A}
}

@INPROCEEDINGS{2023pcsf.conf..308N,
       author = {{Neri}, R. and {S{\'a}nchez-Portal}, M.},
        title = "{IRAM observatories today and tomorrow}",
    booktitle = {Physics and Chemistry of Star Formation: The Dynamical ISM Across Time and Spatial Scales},
         year = 2023,
       editor = {{Ossenkopf-Okada}, V. and {Schaaf}, R. and {Breloy}, I. and {Stutzki}, J.},
        month = feb,
        pages = {308},
       adsurl = {https://ui.adsabs.harvard.edu/abs/2023pcsf.conf..308N}
}

@ARTICLE{2009IEEEP..97.1463W,
       author = {{Wootten}, Alwyn and {Thompson}, A. Richard},
        title = "{The Atacama Large Millimeter/Submillimeter Array}",
      journal = {IEEE Proceedings},
         year = 2009,
        month = aug,
       volume = {97},
       number = {8},
        pages = {1463-1471},
          doi = {10.1109/JPROC.2009.2020572},
archivePrefix = {arXiv},
       eprint = {0904.3739},
 primaryClass = {astro-ph.IM},
       adsurl = {https://ui.adsabs.harvard.edu/abs/2009IEEEP..97.1463W}
}

@INPROCEEDINGS{2023pcsf.conf..304C,
       author = {{Carpenter}, John and {Brogan}, Crystal and {Iono}, Daisuke and {Mroczkowski}, Tony},
        title = "{The ALMA Wideband Sensitivity Upgrade}",
    booktitle = {Physics and Chemistry of Star Formation: The Dynamical ISM Across Time and Spatial Scales},
         year = 2023,
       editor = {{Ossenkopf-Okada}, V. and {Schaaf}, R. and {Breloy}, I. and {Stutzki}, J.},
        month = feb,
        pages = {304},
          doi = {10.48550/arXiv.2211.00195},
archivePrefix = {arXiv},
       eprint = {2211.00195},
 primaryClass = {astro-ph.IM},
       adsurl = {https://ui.adsabs.harvard.edu/abs/2023pcsf.conf..304C}
}

@INPROCEEDINGS{2014SPIE.9153E..1PB,
       author = {{Benson}, B.~A. and {Ade}, P.~A.~R. and {Ahmed}, Z. and {Allen}, S.~W. and {Arnold}, K. and {Austermann}, J.~E. and {Bender}, A.~N. and {Bleem}, L.~E. and {Carlstrom}, J.~E. and {Chang}, C.~L. and {Cho}, H.~M. and {Cliche}, J.~F. and {Crawford}, T.~M. and {Cukierman}, A. and {de Haan}, T. and {Dobbs}, M.~A. and {Dutcher}, D. and {Everett}, W. and {Gilbert}, A. and {Halverson}, N.~W. and {Hanson}, D. and {Harrington}, N.~L. and {Hattori}, K. and {Henning}, J.~W. and {Hilton}, G.~C. and {Holder}, G.~P. and {Holzapfel}, W.~L. and {Irwin}, K.~D. and {Keisler}, R. and {Knox}, L. and {Kubik}, D. and {Kuo}, C.~L. and {Lee}, A.~T. and {Leitch}, E.~M. and {Li}, D. and {McDonald}, M. and {Meyer}, S.~S. and {Montgomery}, J. and {Myers}, M. and {Natoli}, T. and {Nguyen}, H. and {Novosad}, V. and {Padin}, S. and {Pan}, Z. and {Pearson}, J. and {Reichardt}, C. and {Ruhl}, J.~E. and {Saliwanchik}, B.~R. and {Simard}, G. and {Smecher}, G. and {Sayre}, J.~T. and {Shirokoff}, E. and {Stark}, A.~A. and {Story}, K. and {Suzuki}, A. and {Thompson}, K.~L. and {Tucker}, C. and {Vanderlinde}, K. and {Vieira}, J.~D. and {Vikhlinin}, A. and {Wang}, G. and {Yefremenko}, V. and {Yoon}, K.~W.},
        title = "{SPT-3G: a next-generation cosmic microwave background polarization experiment on the South Pole telescope}",
    booktitle = {Millimeter, Submillimeter, and Far-Infrared Detectors and Instrumentation for Astronomy VII},
         year = 2014,
       editor = {{Holland}, Wayne S. and {Zmuidzinas}, Jonas},
       series = {Society of Photo-Optical Instrumentation Engineers (SPIE) Conference Series},
       volume = {9153},
        month = jul,
          eid = {91531P},
        pages = {91531P},
          doi = {10.1117/12.2057305},
archivePrefix = {arXiv},
       eprint = {1407.2973},
 primaryClass = {astro-ph.IM},
       adsurl = {https://ui.adsabs.harvard.edu/abs/2014SPIE.9153E..1PB}
}

@ARTICLE{2023ApJS..264....7C,
       author = {{CCAT-Prime Collaboration}},
        title = "{CCAT-prime Collaboration: Science Goals and Forecasts with Prime-Cam on the Fred Young Submillimeter Telescope}",
      journal = {\apjs},
         year = 2023,
        month = jan,
       volume = {264},
       number = {1},
          eid = {7},
        pages = {7},
          doi = {10.3847/1538-4365/ac9838},
archivePrefix = {arXiv},
       eprint = {2107.10364},
 primaryClass = {astro-ph.CO},
       adsurl = {https://ui.adsabs.harvard.edu/abs/2023ApJS..264....7C}
}

@ARTICLE{2002PASP..114....1W,
       author = {{Weisskopf}, M.~C. and {Brinkman}, B. and {Canizares}, C. and {Garmire}, G. and {Murray}, S. and {Van Speybroeck}, L.~P.},
        title = "{An Overview of the Performance and Scientific Results from the Chandra X-Ray Observatory}",
      journal = {\pasp},
         year = 2002,
        month = jan,
       volume = {114},
       number = {791},
        pages = {1-24},
          doi = {10.1086/338108},
archivePrefix = {arXiv},
       eprint = {astro-ph/0110308},
 primaryClass = {astro-ph},
       adsurl = {https://ui.adsabs.harvard.edu/abs/2002PASP..114....1W}
}

\end{document}